\documentclass[twocolumn, twocolappendix]{aastex631}

\usepackage{amsmath} 
\usepackage{float}
\usepackage{breqn}
\usepackage{threeparttable}
\usepackage{CJKutf8}
\usepackage{multirow}
\definecolor{mygreen}{RGB}{0,128,0}

\begin{document}
\title{Prospects of a New $L_5$ Trojan Flyby Target for the Lucy Mission}

\author[0000-0002-6514-318X]{Luis E. Salazar Manzano}
\affiliation{Department of Astronomy, University of Michigan, Ann Arbor, MI 48109, USA}

\author[0000-0001-6942-2736]{David W. Gerdes}
\affiliation{Department of Physics, Case Western Reserve University, Cleveland, OH 44106, USA}
\affiliation{Department of Physics, University of Michigan, Ann Arbor, MI 48109, USA}
\affiliation{Department of Astronomy, University of Michigan, Ann Arbor, MI 48109, USA}

\author[0000-0003-4827-5049]{Kevin J. Napier} \affiliation{Michigan Institute for Data and AI in Society, University of Michigan, Ann Arbor, MI 48109, USA}
\affiliation{Department of Physics, University of Michigan, Ann Arbor, MI 48109, USA}
\affiliation{Center for Astrophysics $\mid$ Harvard \& Smithsonian, Cambridge, MA 02138, USA}


\author[0000-0001-7737-6784]{Hsing~Wen~Lin}
\affiliation{Department of Physics, University of Michigan, Ann Arbor, MI 48109, USA} \affiliation{Michigan Institute for Data and AI in Society, University of Michigan, Ann Arbor, MI 48109, USA}

\author[0000-0002-8167-1767]{Fred C. Adams}
\affiliation{Department of Physics, University of Michigan, Ann Arbor, MI 48109, USA}

\author[0009-0000-4697-5450]{Tessa Frincke}
\affiliation{Department of Physics, University of Michigan, Ann Arbor, MI 48109, USA}
\affiliation{Department of Physics and Astronomy, University of Toledo, Toledo, OH 43606, USA}
\affiliation{Department of Physics and Astronomy, Michigan State University, East Lansing, MI 48824, USA}

\author[0000-0003-2548-3291]{Simone Marchi}
\affiliation{Southwest Research Institute, Boulder, CO 80302, USA}

\author[0000-0002-6013-9384]{Keith S. Noll}
\affiliation{NASA/Goddard Space Flight Center, Greenbelt, MD 20771, USA}

\author[0000-0003-4452-8109]{John Spencer}
\affiliation{Southwest Research Institute, Boulder, CO 80302, USA}





\begin{abstract}

NASA's Lucy spacecraft is en route to conduct the first close encounter with Jupiter's Trojans. While most scheduled flybys lie in the $L_4$ cloud, the only $L_5$ target is the Patroclus-Menoetius binary. Since each flyby offers unique insights into target and population properties unattainable from Earth, we examine the feasibility of including an additional, yet unknown, $L_5$ target while minimizing the impact on Lucy's primary mission. We use the background $L_5$ Trojans brighter than the completeness limit to model their absolute magnitude, spatial, and orbital distributions. A semi-analytical approach estimates the number of Trojans accessible to Lucy for a given $\Delta v$ budget in both pre- and post-Patroclus scenarios.  Our results indicate that, while it is unlikely that any suitable Trojan lies on Lucy's nominal path, a moderate $\Delta v$ investment ($35-50\,\mathrm{m/s}$) could enable a sub-kilometer ($500-700\,\mathrm{m}$) flyby prior to the Patroclus encounter. Post-Patroclus, the likelihood of a similar flyby is $\sim60\%$ for $\Delta v\sim$ 50 m/s. Simulations with synthetic Trojans reveal that potential targets cluster near the node opposite to the encounter window, producing an optimal search period in late 2026 for both scenarios. Surveying the densest $10\%$ of this region would require under 5 nights with Subaru/HSC or under 2 nights with Rubin, using shift-and-stack techniques. A successful sub-kilometric flyby would expand Lucy's Trojan target size range and provide new constraints on collisional evolution and the long-standing asymmetry in the $L_4/L_5$ clouds. This nodal-clustering strategy could guide target searches in future Lucy extensions or other planetary flyby missions.

\end{abstract}



\section{Introduction} \label{sec:intro}

Flybys of small Solar System bodies by spacecraft provide an unparalleled opportunity to study these primitive objects, whose small sizes and vast distances make them otherwise difficult to characterize from Earth. A close flyby enables disk-resolved imaging and spectroscopy, allowing for detailed characterization of surface features and composition, crucial for reconstructing their thermochemical histories \citep{Emery:2024SSRv}. {\em In situ} observations from multiple viewing angles further enable the determination of three-dimensional shapes and the identification of satellites, binary companions, or circum-body material \citep{Noll:2023SSRv}. Additionally, analysis of surface morphology, such as cratering records, offers insights into the formation and collisional history of these populations \citep{Marchi:2023SSRv}. A key advantage of flybys is the direct determination of an object’s mass through its gravitational influence on the spacecraft’s trajectory, which, when combined with volume estimates from imaging, yields bulk densities and structural properties \citep{Mottola:2024SSRv}.

Despite these significant advantages, only a limited number of small Solar System bodies have been visited by spacecraft. The technological challenges and high costs associated with designing, launching, and operating missions to these objects have constrained their exploration. Most of the stable small body populations have been explored {\em in situ}, including Main Belt asteroids such as Vesta visited by the Dawn spacecraft \citep{Russell:2011SSRv, Russell:2012Sci}, Near Earth asteroids like Bennu visited by OSIRIS-REx \citep{Lauretta:2017SSRv, Dellagiustina:2019NatAs}, comets such as Churyumov–Gerasimenko visited by Rosetta \citep{Glassmeier:2007SSRv, Sierks:2015Sci}, and Kuiper Belt objects such as Arrokoth visited by New Horizons \citep{Stern:2018SSRv, Stern:2019Sci}. Jupiter Trojans remain the most populous major group yet to be visited by a spacecraft. This gap will soon be filled by NASA’s Lucy mission, launched on October 16, 2021, which aims to explore a selection of Jupiter Trojans from both the $L_4$ and $L_5$ clouds \citep{Levison:2021PSJ}. 

Jupiter Trojans are crucial for understanding the early Solar System due to their strong dynamical links to the primordial Kuiper Belt. Modern dynamical models suggest that the original Jupiter Trojan population was lost during the giant planet orbital instability and subsequently replaced by objects scattered inward from the Kuiper Belt when Jupiter reached its current orbit \citep{Nesvorny:2018ARA&A, Bottke:2023PSJ}. If this dynamical model is correct, the Jupiter Trojans are an accessible window of Kuiper Belt material and can provide important clues about the processes that shaped the early Solar System \citep{Bottke:2023SSRv}.

Lucy is an unprecedented mission in terms of small-body exploration, as it will fly by nearly as many asteroids in a single mission as have been visited in decades of prior spacecraft encounters. Its target selection is designed to maximize the diversity of observed Jupiter Trojans \citep{Levison:2021PSJ}. Among its five planned flybys, four will occur in the  $L_4$  cloud, while the only planned  $L_5$  encounter is with the binary system Patroclus-Menoetius on March 3, 2033. Currently, no additional known  $L_5$  Jupiter Trojan is accessible to Lucy along its trajectory, but this does not preclude the existence of an undiscovered Trojan that may be within reach. The current observational completeness limit for Jupiter Trojans corresponds to $\sim\!10\,\mathrm{km}$. The smallest primary mission target for Lucy is Polymele, whose occultation data suggest an elongated shape with axes of 26.2 and 12.8 $\mathrm{km}$ \citep{Buie:2022DPS, Buie:2023LPICo}. Therefore, if it exists, any additional encounterable target would likely be smaller than Lucy's smallest planned Trojan flyby target. If visited, such a target would contribute to the mission's goal of understanding Trojans diversity by extending the sampled size range. Lucy has successfully flown by two MBAs, the smallest of which, Dinkinesh, is only $\!\sim\!700$ m in diameter \citep{Levison:2024Natur}, demonstrating that small Trojans are within the capabilities of the spacecraft operation.

Expanding the $L_5$ Trojans sample could be relevant to understanding the well-established asymmetry between the two clouds. Observations indicate that the leading cloud contains more objects than the trailing cloud, with a ratio of $f_{45}=N_4/N_5\!\sim\!1.4$ \citep{Grav:2011ApJ, Uehata:2022AJ}. This asymmetry has been confirmed to be real and not an artifact of observational biases \citep{Szabo:2007MNRAS}. It can not be attributed to stability conditions, as both clouds exhibit nearly equal escape rates ($\!\sim\!24\%$) over the lifetime of the Solar System \citep{Holt:2020MNRAS}. Proposed explanations include differences in the initial capture process, such as an early perturbation by a transient fifth ice giant during planetary migration \citep{Nesvorny2013ApJ}. Alternatively, models where Trojans are captured from the core feeding zone during a Jupiter inward migration can produce asymmetric clouds, though they fail to match observational constraints on total mass and inclination distribution \citep{Pirani:2019A&Aa, Pirani:2019A&Ab}. More recent observations suggest the asymmetry may be linked to albedo and/or shape irregularities at the small-size scales \citep{Vokrouhlicky:2024AJ, McNeill:2021PSJ}. Increasing the number of $L_5$ flyby targets would enable more robust comparisons of physical properties between the two clouds, potentially shedding light on this long-standing problem. 

To date, nearly all spacecraft flyby targets have been known objects at the time of mission launch, with the exception of Arrokoth. Before the launch of New Horizons, its long flight time and the lack of other planned missions to explore the outer Solar System prompted studies on the feasibility of identifying a post-Pluto flyby target \citep{Spencer:2003EM&P}. The discovery of Arrokoth required a large-scale, multi-year observational campaign using both ground- and space-based telescopes, but it was ultimately identified using the Hubble Space Telescope \citep{Spencer:2015EPSC, Buie:2024PSJ}. This discovery was invaluable, for example, Arrokoth's contact-binary structure supports the pebble cloud collapse formation model over one involving high-velocity impacts \citep{Stern:2019Sci}, and its lower-than-expected crater density suggests that the Cold Classical Kuiper Belt has been less dynamically active than previously assumed \citep{Singer:2019Sci}.

Motivated by the rarity of flyby opportunities and their potential to unlock new insights into Solar System science, this work investigates the feasibility of identifying an additional, yet unknown,  $L_5$  Jupiter Trojan flyby target for Lucy. In Section \ref{sec:model}, we describe our modeling of the $L_5$ cloud, including its absolute magnitude distribution, spatial structure, and key orbital elements. Section \ref{sec:probability} defines the timing and geometry of potential flyby opportunities and presents our estimates for finding an additional target. Section \ref{sec:search} discusses the observational requirements and constraints for detecting such a target. Finally, Section \ref{sec:discussion} expands on the implications of our results, followed by a summary and conclusions in Section \ref{sec:conclusions}. Lucy and Jupiter state vectors were retrieved using SPICE kernels, and all the n-body integrations described in this work were performed using the Python package \texttt{spacerocks} \citep{Napier:2020spacerocks}.

\section{Model of the $L_5$ Cloud} \label{sec:model}

Examining the possibility of finding a new flyby target requires a focus on physical and orbital regimes that remain poorly characterized. This presents a significant challenge as any predictions must be robust enough to justify the mobilization of resources. To address this, we leverage the latest constraints on observational completeness limits and collisional families.

Recent studies have determined that below the absolute magnitude limit of $H_V = 13.8$ ($D\!\sim\!10$ km) no additional Trojans have been identified and added to the list of known objects \citep{Vokrouhlicky:2024AJ, Hendler:2020PSJ}. This corresponds to an $r$-band absolute magnitude completeness limit of $H_r = 13.55$ assuming a $V-r$ color of 0.25 mag \citep{Szabo:2007MNRAS}. This would be useful as we can use the population of $L_5$ Jupiter Trojans brighter than this limit to determine the distribution of orbital properties that do not depend on size. \cite{Vokrouhlicky:2024AJ} also presents the most recent separation of the background continuous population of Trojans from the discrete members of the collisional families. This two-component architecture of the Jupiter Trojan cloud is important to avoid biases due to overdensities in the regions of the parameter space where collisional families are clustered. At the time of writing, there are 5467 $L_5$ Jupiter Trojans listed in the Minor Planet Center Orbit Database (MPCORB). By eliminating the members in the 4 $L_5$ collisional families identified in \cite{Vokrouhlicky:2024AJ}, a total of 5002 are part of the background population. Finally, by considering only those Trojans brighter than the completeness limit, we will model the $L_5$ cloud with a sample of 1404 Jupiter Trojans.

\subsection{Absolute magnitude distribution} \label{subsec:Hdistr}

\begin{figure}
    \centering
    \includegraphics[width=1\linewidth]{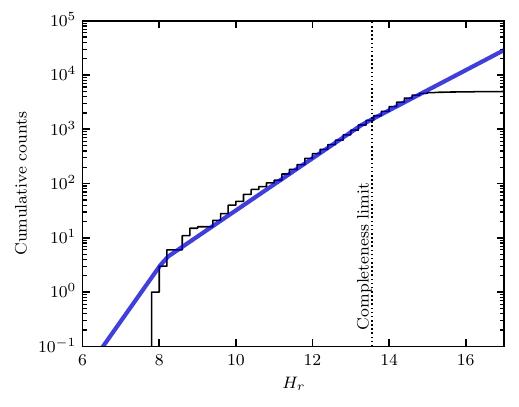}
    \caption{$H_r$ distribution model of $L_5$ Jupiter Trojans. The black line represents the cumulative histogram of MPC-listed $L_5$ Jupiter Trojans of the background population. The blue line shows the model adopted in this work. The black vertical dotted line indicates the observational completeness limit.}
    \label{fig:Hdistr}
\end{figure}

The $H$ distribution has been constrained for both Jupiter Trojan clouds down to $H_r \!\sim\! 17$ ($D \!\sim\! 2\,\mathrm{km}$), with the most stringent constraints provided by the Hyper Suprime-Cam on the $8.2\,\mathrm{m}$ Subaru Telescope \citep{Uehata:2022AJ, Yoshida:2017AJ}. The absolute magnitude distribution of Jupiter Trojans is typically modeled as a broken power law. However, the precise locations of the breaks and the slopes in the different regimes remain topics of debate, as observational constraints vary between studies (e.g. \citealt{Wong:2015AJ, Yoshida:2005AJ}).

For the purposes of this work, we model the cumulative absolute magnitude distribution of the L$_5$ Jupiter Trojans using a power law with two breaks, as described in Equation \eqref{eq:hdistr}:


\begin{dmath}
\Sigma(<H) =
\begin{cases}
10^{\alpha_1 (H - H_0)} \quad \text{if } H < H_{B1}, \\
10^{\alpha_2 (H - H_{B1}) + \alpha_1 (H_{B1} - H_0)} \quad \text{if } H_{B1} \leq H < H_{B2}, \\
10^{\alpha_3 (H - H_{B2}) + \alpha_2 (H_{B2} - H_{B1}) + \alpha_1 (H_{B1} - H_0)} \quad \text{if } H \geq H_{B2}.
\end{cases}
\label{eq:hdistr}
\end{dmath}

\begin{figure*}
    \centering
    \includegraphics[width=1\textwidth]{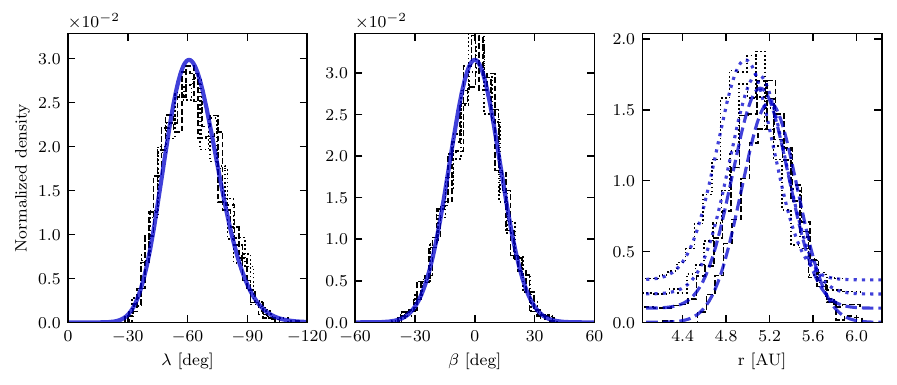}
    \caption{Spatial distribution model of the $L_5$ Jupiter Trojans in a coordinate system co-rotating with Jupiter. The first, second, and third panels represent, respectively, the longitude ($\lambda$), latitude ($\beta$), and radial distance ($r$) distributions. The histograms correspond to MPC background Trojans brighter than the completeness limit for four different epochs, and the blue lines the fitted models. The radial distribution includes a vertical offset for visual clarity.}
    \label{fig:spatialdistr}
\end{figure*}

Here, $H_0$ determines the normalization of the distribution while the remaining parameters its shape. In particular, $H_{B1}$ and $H_{B2}$ are the locations of the two breaks, and $\alpha_1$, $\alpha_2$, and $\alpha_3$ are the slopes in the three regimes. For the large-size regime, we adopt $H_{B1} = 8.4$ ($H_{B1} = 8.15$ in the $r$ band) and $\alpha_1 = 1$, as determined by \cite{Fraser:2014ApJ}. The values $\alpha_2 = 0.46$ and $\alpha_3 = 0.37$ are taken from \cite{Uehata:2022AJ}, while $H_{B2} = 13.56$ ($H_{B2} = 13.31$ in the $r$ band) is adopted from \cite{Yoshida:2017AJ}. The normalization factor $H_0$ was determined by fitting Equation \eqref{eq:hdistr} to the cumulative histogram of the absolute magnitudes of background MPC-listed $L_5$ Trojans brighter than the completeness limit (see Figure \ref{fig:Hdistr}). Given that the typical precision of magnitudes reported to the MPC is 0.1 mag \citep{Hendler:2020PSJ}, we used 0.2 mag bins and excluded the last bin when fitting the cumulative histogram.

The absolute magnitude distribution serves as a proxy for the size distribution, under the assumption of a uniform albedo for Jupiter Trojans. WISE thermal measurements report a typical albedo of $p\sim$~0.05 independent of size \citep{Grav:2011ApJ, Grav:2012ApJ, Romanishin:2018AJ}. However, \cite{Fernandez:2009AJ} suggest that smaller Trojan may exhibit higher albedos, perhaps due to collisional resurfacing. \cite{Simpson:2022AJ} confirmed elevated albedos for some Trojans, and recent JWST observations of a subsample indicate that they may belong to a new taxonomic class \citep{Brown:2025PSJ}. It remains unclear whether this trend is intrinsic or driven by observational biases and uncertainties \citep{Emery:2024SSRv}. Because these uncertainties do not affect the underlying $H$ distribution but do propagate into diameter estimates, we retain the conventional choice of $p=0.05$ and revisit the implications of albedo distribution and diameter conversion in Section~\ref{subsec:hsensi}.

\subsection{Spatial distribution} \label{subsec:spatialdistr}

The $L_5$ Jupiter Trojans are confined within a volume around the stable Lagrange point located $60^\circ$ behind Jupiter in its orbit \citep{Levison:1997Natur}. The most practical coordinate system for describing their spatial distribution is then a co-rotating frame with Jupiter. We adopt the Jupiter Solar Orbital (JSO) coordinate system \citep{Wang:2023E&SS}, where the $x$-axis points from the Sun to Jupiter\footnote{In the original JSO system, the $x$-axis has the origin in Jupiter.}, the $z$-axis is perpendicular to Jupiter’s orbital plane in the northward direction, and the $y$-axis completes the right-handed system, positive in the direction of Jupiter's motion. The rotation matrix $\mathbf{R}(t)$ transforms the inertial position $\mathbf{r}(t)$ into the co-rotating frame, giving:

\begin{equation}
    \mathbf{r}'(t) = \mathbf{R}(t) \, \mathbf{r}(t)
    \label{eq:rot}
\end{equation}

We assume that the spatial distribution of $L_5$ Jupiter Trojans is independent of size. This follows from the collisional communication within the cloud, which homogenizes the size distribution across all orbital phase space \citep{Vokrouhlicky:2024AJ}. Consequently, the distribution of Trojans brighter than the completeness limit serves as a representative approximation of the entire $L_5$ Trojan population. To mitigate correlations between cartesian coordinates in the co-rotating frame, we describe the distribution in spherical coordinates $(\lambda,\,\beta,\,r)$, where $\lambda$ represents the longitude, $\beta$ the latitude, and $r$ the heliocentric distance. We model their probability distributions, $f_\lambda$, $f_\beta$, and $f_r$, independently.

The $f_\lambda\,f_\beta$ distribution in Jupiter's coordinate system has typically been modeled as a bivariate normal distribution (e.g., \citealt{Szabo:2007MNRAS}). Here we use for $f_\lambda$ a lognormal distribution ($ f(x |\mu,s,\sigma) = \frac{1}{(x-\mu) \sigma \sqrt{2\pi}} \,\exp\left( -\frac{\ln^2(( x - \mu)/s)}{2\sigma^2} \right)$) as it better accounts for the known asymmetry in the librational motion around the equilibrium point that occurs with the increase in the amplitude of the libration for tadpole orbits \citep{Marzari:2002aste.book}. For both $f_\beta$ and $f_r$ we used a Gaussian distribution ($ f(x |\mu,\sigma) = \frac{1}{\sigma \sqrt{2\pi}} \,\exp\left( -\frac{(x-\mu)^2}{2\sigma^2} \right)$). To determine the model parameters we fit these functions to the coordinates of background Trojans brighter than the completeness limit using a Maximum Likelihood Estimation method. We propagate the Trojans orbits over a $\sim\!20$ year time span, beginning five years prior to Lucy's launch and extending to five years beyond the Patroclus encounter. 

The fitting of $f_\lambda$ was performed by fixing the location parameter to $\mu_{\lambda}=-60^\circ$, and we use the resulting average of the best-fit values for the shape parameter $\sigma_\lambda$ and scale parameter $s_\lambda$. These two parameters exhibit variations of $\!\sim\!1\,\%$ throughout the integration window, and are thus treated as time independent. For $f_\beta$ we fit the Gaussian fixed to $\mu_\beta=0^\circ$, and found the fitted standard deviation $\sigma_\beta$ to be similarly stable over time. In contrast, the heliocentric distance of $L_5$ Trojans varies throughout their orbit between perihelion and aphelion. As a result, we allowed both $\mu_r$ and $\sigma_r$ in the $f_r$ distribution to vary. We will use the average value for the standard deviation $\sigma_r$ as only shows minor variation ( $\!\sim\!1\,\%$). The mean, however, was modeled with a sinusoidal representation $\mu_r(t)=A_r \sin(2\pi\, t /P_r+ \phi_r) + C_r$. The final fitted parameters for all distributions are summarized in Table \ref{tab:trojan_parameters}. Figure \ref{fig:spatialdistr} presents an example of the data used for the fits and the corresponding models for four different epochs. Histograms were generated using Scott’s rule \citep{Scott:1979} for visualization only.

\begin{table}
\centering
\caption{Fitted parameters for the spatial distribution of the $L_5$ background Trojan population.}
\label{tab:trojan_parameters}
\setlength{\tabcolsep}{2pt} 
\begin{tabular}{cccc}
\hline
\hline
\textbf{Distribution} & \textbf{Model} & \textbf{Parameter} & \textbf{Value} \\
\hline
\multirow{3}{*}{Longitude ($\lambda$)} & \multirow{3}{*}{Lognormal} 
    & $\sigma_\lambda$ & 0.11 \\
& & $s_\lambda$ & 122.2$^\circ$ \\
& & $\mu_\lambda$ & $-60^\circ$ \\
\hline
\multirow{2}{*}{Latitude ($\beta$)} & \multirow{2}{*}{Gaussian} 
    & $\sigma_\beta$ & 12.64$^\circ$ \\
& & $\mu_\beta$ & $0^\circ$ \\
\hline
\multirow{6}{*}{Heliocentric distance ($r$)} & \multirow{6}{*}{Gaussian} 
    & $\sigma_r$ & 0.26 AU \\
& & $\mu_r(t)$ & - \\
& & $A_r$ & 0.24 AU \\
& & $P_r$ & 4316.06 d \\
& & $\phi_r$ & 4.85 rad \\
& & $C_r$ & 5.21 AU \\
\hline
\hline
\end{tabular}
\end{table}

The spatial distributions are treated as probability density functions (PDFs), therefore they are normalized to satisfy the following condition for a fixed time $t$:


\begin{equation}
\int_{0}^{\infty} \int_{-90^\circ}^{90^\circ} \int_{-180^\circ}^{180^\circ} f_\lambda(\lambda) f_\beta(\beta) f_r(r|\,t) \, d\lambda \, d\beta \, dr = 1
\end{equation}

meaning the full joint probability distribution function is given by:





\begin{equation}
f_\lambda(\lambda) f_\beta(\beta) f_r(r|\,t) = - \frac{e^{\gamma(\lambda,\,\beta,\,r|\,t)}}{( \lambda+\mu_{\lambda} ) \sigma_\lambda \sigma_\beta \sigma_r (2\pi)^{3/2}}  
,
\label{eq:pdf}
\end{equation}

where 
\begin{equation}
\gamma(\lambda,\beta,r|\,t)\!=\!
-\frac{\ln^2 ( -\!\left(  \lambda \!+ \!\mu_\lambda \right)\!/\!s_{\lambda} ) }{2 \sigma_\lambda^2}
 -\frac{(\beta\! - \!\mu_\beta)^2}{2 \sigma_\beta^2} 
-\frac{(r \!- \!\mu_r\!\left(t\right))^2}{2 \sigma_r^2}
.
\end{equation}

Note that a factor of ($-1$) has been applied to $\lambda$ to ensure compatibility with the lognormal distribution, which is only defined for positive values. This accounts for the fact that, in our coordinate system, $L_5$ is located within the $0^\circ>\lambda>-180^\circ$ hemisphere. The joint PDF, $f_\lambda f_\beta f_r$, represents the normalized spatial density $n_{\text{sph}}$ at a spherical location ($\lambda, \beta, r$):


\begin{equation}
n_{\text{sph}}(\lambda, \beta, r|\,t)=f_\lambda(\lambda) f_\beta(\beta) f_r(r|\,t) 
\label{eq:nsph}
\end{equation}

This expression describes the density per solid angle per unit distance. To express it as a normalized number density in Cartesian coordinates ($x$, $y$, $z$), which enables the use of convenient units like $\mathrm{AU^3}$, we apply the appropriate coordinate transformation:


\begin{equation}
n(x,y,z|\,t)=
f_\lambda(\lambda(\mathbf{r}')) 
f_\beta(\beta(\mathbf{r}')) 
f_r(r|\,t)
\left| \frac{\partial (\lambda, \beta, r)}{\partial (x, y, z)} \right| 
\label{eq:nnorm}
\end{equation}

where the Jacobian of the transformation is:

\begin{equation}
\left| \frac{\partial (\lambda, \beta, r)}{\partial (x, y, z)} \right| 
= \frac{1}{\sqrt{x^2 + y^2 + z^2} \, \sqrt{x^2 + y^2}}=\frac{1}{r^2 cos\beta}
.
\label{eq:jacobian}
\end{equation}

Note that \cite{Nakamura:2008PASJ} also considered a longitudinally elongated distribution, specifically proposing a Maxwellian-Gaussian model for the longitudinal distribution. However, their approach was based on a surface density model observed from Earth, limited to opposition and described in ecliptic coordinates. In contrast, our spatial distribution model formulated in a Jupiter co-rotating frame, reduces the number of free parameters, and can be transformed into an observed surface density that can be evaluated at any arbitrary phase point.  

Finally, by multiplying the absolute magnitude distribution obtained in Section \ref{subsec:Hdistr} with the normalized spatial distribution in Equation \eqref{eq:nnorm}, we obtain the number density of $L_5$ Jupiter Trojans larger than a given absolute magnitude at any point in Jupiter's corotating frame:

\begin{equation}
n_H(x,y,z, H|\,t)=\Sigma(<H) \, n(x,y,z|\,t)
\end{equation}

\subsection{Synthetic Trojans} \label{subsec:fakes}

To generate a population of synthetic objects for exploring the encounter probability (Section \ref{subsec:integral}) and the search conditions (Section \ref{sec:search}), we adopt the spherical orbit parametrization introduced by \cite{NapierHolman:2024PSJ}. This new orbital basis is ideally suited for our study, as it allows initializing synthetic objects using their location in spherical coordinates $\{\varphi, \theta, r\}$ at a given epoch. The remaining part of its orbital state is specified by defining Keplerian elements $\{a, e, i\}$ consistent with $L_5$ Jupiter Trojans.  

A detailed modeling of the orbital element distribution is beyond the scope of this work. However, analogous to Section \ref{sec:model}, we approximate the distribution of known $L_5$ Trojans from the background population down to the completeness limit. Specifically, we model the semi-major axis $a$ as a Gaussian distribution centered at $5.2\,\mathrm{AU}$. The orbital inclination $i$ and eccentricity $e$ are modeled using Rayleigh distributions  ($ f(x|\, \sigma) = \frac{x}{\sigma^2}\,\exp\left( -\frac{x^2}{2\sigma^2}\right)$) for $x\geq0$.


The models resulting from our fits are presented in the appendix as Figure \ref{fig:orbitaldistr}. Excluding collisional family members and focusing on the background population is particularly important when modeling the inclination distribution. As noted by \cite{Vokrouhlicky:2024AJ}, the well-known bimodality in $i$ \citep{Jewitt:2000AJ, Slyusarev:2014SoSyR} is likely due to clustering caused by collisional families.

To ensure the stability and true membership of synthetic Trojans in the  $L_5$  cloud, we integrate their orbits for 1 $\mathrm{Myr}$ with 30-$\mathrm{d}$ time steps, starting from Lucy’s launch date, and including all giant planets of the Solar System. We remove synthetic objects whose semi-major axis evolves outside the range  $5.0 \leq a \leq 5.4\,\mathrm{AU}$ at any point in the simulation. Additionally, we eliminate horseshoe orbits by rejecting objects whose mean longitude relative to Jupiter exceeds $180^\circ$ at any time during the integration. We find that $\!\sim\!99\%$ of the generated synthetic Trojans survive this process, indicating that the original spatial distribution is preserved.

\section{Encounter Probability} \label{sec:probability}

\subsection{Encounter windows} \label{subsec:windows}

\begin{figure}
    
    \centering
    \begin{minipage}{\linewidth}
        \centering
        \includegraphics[width=\linewidth]{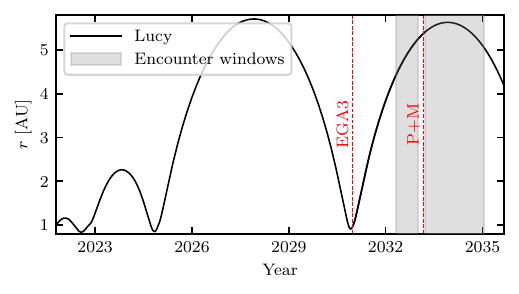}
    \end{minipage}
    
    
    \centering
    \begin{minipage}{\linewidth}
        \centering
        \includegraphics[width=\linewidth]{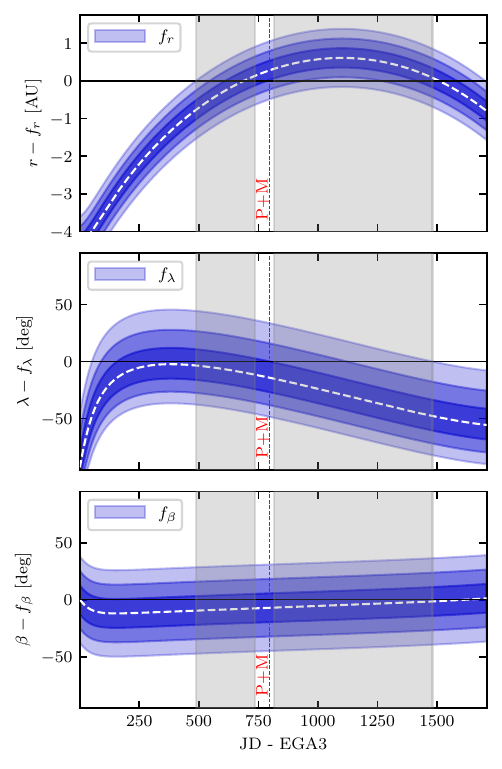}   
    \end{minipage}

   \caption{(\textbf{top}) Heliocentric distance of Lucy over time. Red vertical lines indicate the third Earth Gravity Assist (EGA3) and the Patroclus-Menoetius flyby (P+M). Gray shaded regions mark the pre- and post-Patroclus encounter windows explored in this study. (\textbf{bottom}) Lucy's position relative to the $L_5$ cloud distributions ($f_r$, $f_\lambda$, and $f_\beta$) after EGA3. The white dashed line show the relative position of $\mu_r(t)$, $\mu_\lambda$, and $\mu_\beta$, while the shaded blue regions the 1, 2, and 3 $\sigma$ boundaries. The encounter windows are defined by Lucy’s entry and exit relative to the spatial extent of the $L_5$ cloud, as well as by operational constraints associated with Patroclus encounter.}\label{fig:windows}
\end{figure}

If an additional flyby target were to be included in the Lucy mission, its impact on the nominal mission schedule must be minimized. The only scheduled mission event that constrains the timing of a new flyby in the $L_5$ cloud is the Patroclus-Menoetius encounter. Therefore, we define two potential scenarios: a pre-Patroclus encounter window and a post-Patroclus encounter window.

The pre-Patroclus encounter window begins when Lucy enters the inner edge of the $L_5$ cloud. Following Lucy’s third Earth Gravity Assist (EGA3) on December 27, 2030, the spacecraft is directed toward $L_5$, and therefore the cloud ingress is defined only by the radial extent of the $f_r$ distribution. We adopt the $3\sigma_r$ boundary as the effective edge of the cloud and account for the time evolution of $\mu_r(t)$ (Section \ref{subsec:spatialdistr}), which shifts to smaller heliocentric distances as Lucy approaches. Lucy reaches this limit at 4.41 AU on April 27, 2032 (see second panel of Figure \ref{fig:windows}). The key baseline activities common across all Lucy encounters include optical navigation, which begins 60 days before close approach and science observations which conclude 20 days after \citep{Olkin:2024SSRv}. To ensure operational readiness, we define the end of the pre-encounter window as January 2, 2033, 60 days before the Patroclus flyby, when Lucy is at 5.25 AU.

The Lucy mission nominally concludes $\sim$30 days after the Patroclus encounter, therefore, for a post-Patroclus scenario, we assume Lucy remains on the same orbit established after EGA3. We define the start of the post-encounter window as 20 days after the Patroclus flyby, when encounter operations have concluded, in March 23, 2033, with Lucy at 5.41 AU. Although Lucy remains within the radial extent of the $L_5$ cloud, it approaches the trailing edge of the $f_\lambda$ distribution. Since $f_\lambda$ is modeled as a lognormal distribution, we define the cloud's edge by the location enclosing 99.73\% of this distribution. Lucy exits this region on January 16, 2035, at a heliocentric distance of 5.06 AU (see third panel of Figure \ref{fig:windows}).

The resulting pre- and post-Patroclus encounter windows spans $\sim$250 and $\sim$650 days, respectively. These windows are visualized in Figure \ref{fig:windows}, which shows the evolution of Lucy's heliocentric distance (top) and its relative position with respect to the $f_r$, $f_\lambda$, and $f_\beta$ distributions of the $L_5$ cloud (bottom). The four epochs shown in Figures \ref{fig:spatialdistr} and \ref{fig:orbitaldistr} correspond to the start and end times of both windows, dashed lines mark the pre-Patroclus window, and dotted lines the post-Patroclus window.

\subsection{Accessible volume} \label{subsec:volume}

\begin{figure}
    \centering
    \includegraphics[width=1\linewidth]{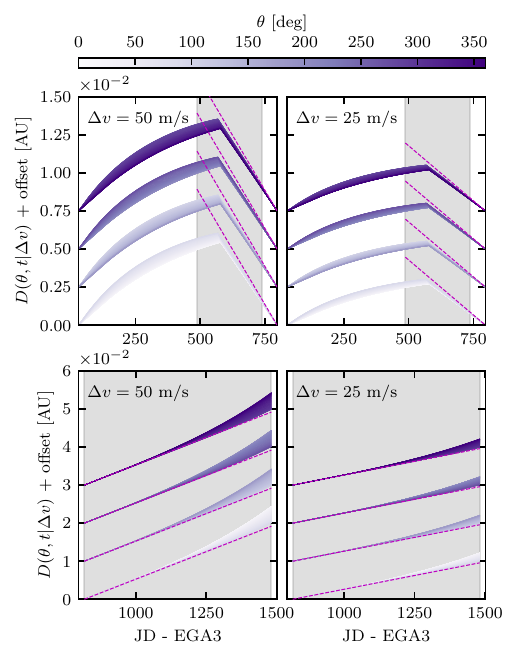}
    \caption{
    Time evolution of the spatial separation between Lucy’s nominal trajectory and trajectories modified with a given $\Delta v$ budget during the pre- (top) and post-Patroclus (bottom) encounter windows. The shaded gray regions mark the defined encounter windows. Left and right columns correspond to $\Delta v = 50\, \mathrm{m/s}$ and $\Delta v = 25\, \mathrm{m/s}$, respectively. Each colored curve represents a different azimuthal direction $\theta$, grouped into 90° bins and vertically offset for clarity. Dashed curves show the linear approximation.
    }
    \label{fig:Devol}
\end{figure}

The space around Lucy’s trajectory during the encounter windows defines the volume where potential $L_5$ Trojan flybys are feasible. A convenient way to describe this space is by considering slices perpendicular to Lucy’s velocity vector at each time (see Appendix \ref{app:cylcoords}). If $\theta$ is the angle in the cross sectional plane, the radius of the cross section is:

\begin{equation}
\rho(\theta,\,t\,|\,\Delta v)=d+D(\theta,\,t\,|\,\Delta v)
\label{eq:rho}
\end{equation}

Here $d$ is the maximum close-approach distance. While Lucy’s planned flybys typically approach within $\sim$1000 km \citep{Olkin:2024SSRv}, we adopt a smaller value of $d \!\simeq\! 500 \, \mathrm{km}$, the distance used for the flyby of the sub-kilometer MBA Dinkinesh \citep{Levison:2024Natur}.

The term $D(\theta, t | \Delta v)$ captures the potential trajectory deviation enabled by a $\Delta v$ budget. A maneuver applied at time $t_m$ broadens the accessible region into a cone-shaped volume. In the absence of maneuvers ($\Delta v = 0$), the accessible space is a cylinder aligned with Lucy’s nominal path. Our focus is on identifying additional low-cost flyby targets with minimal impact on Lucy’s primary schedule. We limit our study to $\Delta v \leq 50\,\mathrm{m/s}$, significantly lower than most planned mission maneuvers\footnote{Most Lucy deep space maneuvers are larger than 100 m/s \citep{Stanbridge:2017}.}.

For pre-Patroclus opportunities, the maneuvering space is constrained by EGA3 and the fixed Patroclus flyby. Two maneuvers are needed, one to depart the nominal trajectory and another to return for the Patroclus encounter. However, the first maneuver may be obtained “for free” at EGA3 by adjusting the geometry of the gravity assist, thus only the return maneuver cost actual fuel. The resulting accessible volume has a “double-cone” shape (with one apex at EGA3 and the other at the Patroclus encounter), and depends on the azimuthal angle $\theta$, polar angle of the initial maneuver, $\Delta v$ budget, and the time of return. We obtain $D(\theta, t | \Delta v)$ through a trajectory optimization algorithm maximizing the accessible volume weighted by the spatial density of $L_5$ Trojans (see Appendix \ref{app:optimization}).

For the post-Patroclus case, the volume is less constrained and obtained with a single maneuver at the start of the window. In this case, the function $D(\theta, t | \Delta v)$ defines a simpler, single-cone volume and is again determined via optimization (see Appendix \ref{app:optimization}).

Figure \ref{fig:Devol} presents the resulting $D(\theta, t | \Delta v)$ profiles for two $\Delta v$ budgets, both before and after the Patroclus encounter. For comparison, a linear approximation $D(t | \Delta v) = |t - t_m| \Delta v $ is shown, which is only valid near the maneuver epoch. The optimal return time in the pre-Patroclus case is $\!\sim\!85$ days after the start of the window.

\subsection{Expected number of Trojans: Semi-analytical approach} \label{subsec:integral}

\begin{figure*}
    
    \centering
    \begin{minipage}{\textwidth}
        \centering
        \includegraphics[width=\linewidth]{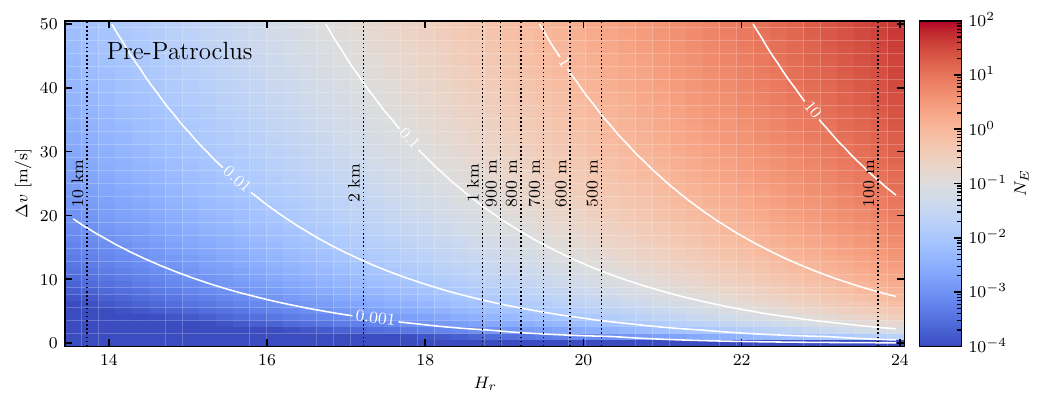}
    \end{minipage}
    
    \centering
    \begin{minipage}{\textwidth}
        \centering
        \includegraphics[width=\linewidth]{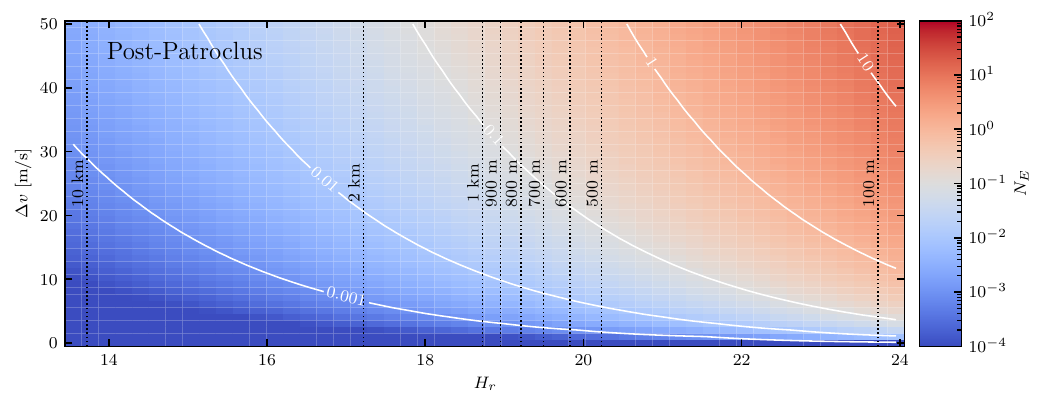}   
    \end{minipage}

    \caption{Expected number of $L_5$ Jupiter Trojans within the volume accessible to Lucy during the pre-Patroclus (top) and post-Patroclus (bottom) encounter windows. White contours indicate values of $N_E = 0.001$, 0.01, 0.1, 1, and 10. Vertical black dotted lines show equivalent absolute magnitudes $H_r$ for different object diameters, assuming a 5\% albedo. While the likelihood of a new flyby target along the nominal trajectory is very low, moderate $\Delta v$ maneuvers can enable additional sub-kilometer flybys, particularly during the pre-Patroclus window.}
\label{fig:NE}
\end{figure*}

The number of $L_5$ Trojans expected within Lucy’s accessible volume during the pre- and post-Patroclus windows (Section~\ref{subsec:volume}) is determined by the local number density of Trojans (Section~\ref{subsec:spatialdistr}) and the limiting object size, characterized by the absolute magnitude $H$ (Section~\ref{subsec:Hdistr}). Formally, this is expressed as a volume integral of the joint spatial-$H$ distribution. Owing to the geometry of the problem, the volume integral can be reformulated as a path integral along Lucy’s trajectory, with the cross-sectional area at each time acting as the integration surface. Given that Lucy’s position and velocity are known as a function of time, the path element is written as $ds = \|\mathbf{v}(t)\| dt$, and the resulting expression for the total expected number of encounterable $L_5$ Trojans as a function of $H$ and $\Delta v$ is:

\begin{equation}
\begin{split}
N_E(H\,|\, \Delta v) = \Sigma(<\!H)\! \,\int_{t_i}^{t_f}\!\!\int_{0}^{2\pi}\!\!&\int_{0}^{\rho(\theta,\,t\, |\, \Delta v)}\!\!n(\mathbf{r}'(\rho,\,\theta,\,t)|\,t)\\
\times \rho \,  d\rho  \,  d\theta & \, \|\mathbf{v}(t)\|dt
\end{split}
\label{eq:NE}
\end{equation}

Here, $\Sigma(<\!H)$ is the cumulative magnitude distribution (Equation~\eqref{eq:hdistr}), $n(\mathbf{r}')$ is the spatial density of Trojans in the co-rotating frame (Equation~\eqref{eq:nnorm}), and the full coordinate transformation is given in Equations~\eqref{eq:rn} and~\eqref{eq:rot}. The radius of the cross section $\rho(\theta,t|\Delta v)$ is given in Equation~\eqref{eq:rho}, and the integration is performed over the time window $[t_i, t_f]$ defined in Section~\ref{subsec:windows}.

The result of numerically evaluating Equation~\eqref{eq:NE} for the pre- and post-Patroclus encounter windows is shown in Figure~\ref{fig:NE}. For a given $N_E$, the relation between $\Delta v$ and magnitude $H_r$ approximately follows a power law. Our results indicate that the probability of finding an additional target along the nominal trajectory (i.e., without maneuvers) is extremely low, on the order of 1 in 100{,}000 for objects as small as 100~m ($H_r \!\sim\! 24$). However, even modest $\Delta v$ increases the likelihood. For example, a $\!\sim\!10\%$ chance of encountering a 500–1000~m object is possible with $\Delta v \!\sim\! 10$–$20\, \mathrm{m/s}$ pre-Patroclus, and $\Delta v \!\sim\! 20$–$35\, \mathrm{m/s}$ post-Patroclus. 

In the pre-Patroclus window, Lucy traverses the core of the $L_5$ cloud, resulting in higher encounter probabilities even for smaller volumes. For instance, with a $\Delta v \!\sim\! 50\, \mathrm{m/s}$, Lucy could expect to encounter a $\!\sim\!700$~m object ($H_r \!\sim\! 19.5$). Targets as small as 600–500~m become accessible with $\Delta v \sim 40$–$35\, \mathrm{m/s}$. In contrast, although the accessible volume post-Patroclus is larger, the local Trojan density is lower, reducing the likelihood of a successful flyby. For example, a $\Delta v\sim$50 m/s yields probabilities of $\sim\!40\%$, $\sim\!55\%$, and $\sim\!75\%$ for accessing a single Trojan of 700 m, 600 m, and 500 m, respectively. A key advantage of the post-Patroclus window, however, lies in its more relaxed operational constraints.

To validate our semi-analytical estimate, we also compute the expected number of Trojans using a Monte Carlo approach. This involves generating synthetic realizations of the $L_5$ Trojan cloud and simulating encounter statistics. Results from this method for the pre-Patroclus case are described in Appendix~\ref{app:numerical} and confirm the accuracy of our semi-analytical framework.




\section{Search Conditions} \label{sec:search}

The favorable probability we found of encountering at least one $L_5$ Trojan within Lucy’s accessible volume, given a moderate $\Delta v$ investment, provides strong motivation to consider a dedicated search effort. However, the decision to proceed depends on several factors, including the specific search conditions and the resources required to conduct it. Although we focus primarily on the pre-Patroclus case, which offers the most favorable probability of detecting an additional target, we also describe the search conditions relevant to the post-Patroclus scenario.

The search conditions are governed by the dynamical properties of $L_5$ Jupiter Trojans that fall within Lucy’s accessible volume. To explore these conditions, we generate a population of synthetic Trojans using a modified version of the procedure presented in Section 4 of \cite{NapierHolman:2024PSJ}. The first key modification is that, rather than injecting objects solely along Lucy's nominal trajectory, we sample synthetic Trojans throughout the accessible volume (see Section~\ref{subsec:volume}). Each object is assigned a time $t$ drawn from the encounter window, weighted by the spatial Trojan density $n(\mathbf{r}'(t))$ along Lucy’s trajectory (Equation~\eqref{eq:nnorm}) within the window. Given this time, the corresponding position $\mathbf{r}(t)$ of Lucy is determined, which combined with a randomly sampled $\theta$ (uniform between 0 and $2\pi$) and $\rho$ (uniform between 0 and the local radius), the object's location around Lucy's path is specified. 

A second key modification is the adoption of realistic orbital element distributions. Specifically, the orbital elements $\{a, e, i\}$ are sampled according to the fits derived in Section~\ref{subsec:fakes}. With a known spatial position in spherical coordinates at the given encounter epoch, these elements fully specify the state vector of the synthetic $L_5$ Trojans. Additionally, we filter out non-tadpole orbits by performing the integration also described in Section \ref{subsec:fakes}. We use a final population of $\sim$10,000 synthetic Jupiter Trojans.

\subsection{When and where to search?} \label{subsec:whenwhere}

\begin{figure*}
    
    \centering
    \begin{minipage}{\textwidth}
        \centering
        \includegraphics[width=\linewidth]{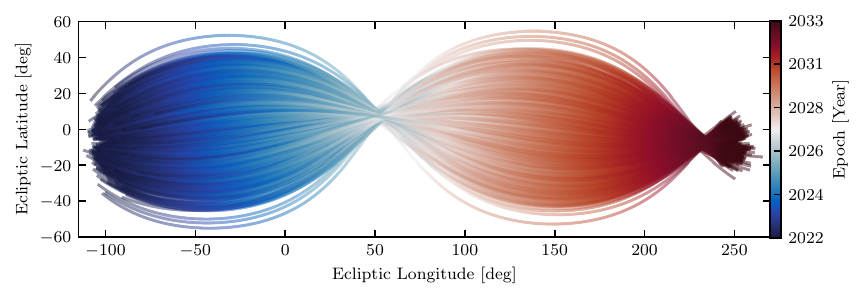}
    \end{minipage}
    
    
    \centering
    \begin{minipage}{\textwidth}
        \centering
        \includegraphics[width=\linewidth]{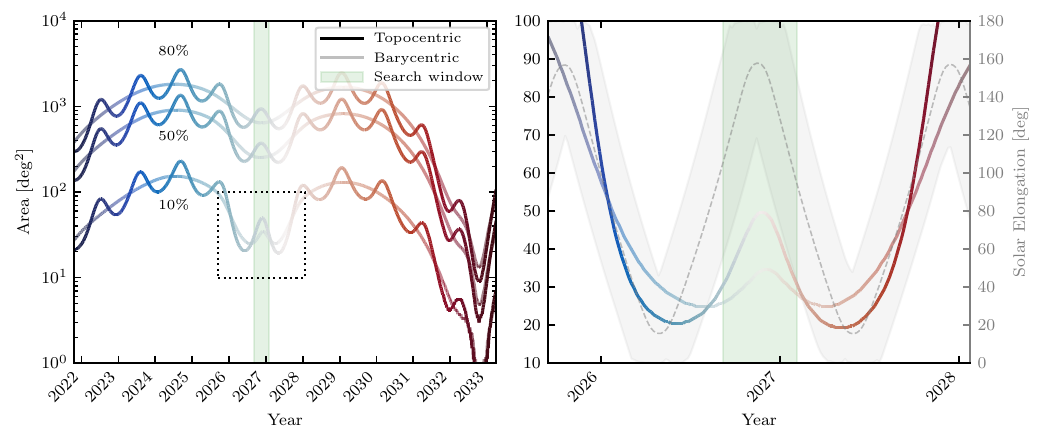}   
    \end{minipage}

   \caption{(\textbf{Top}) Ecliptic barycentric coordinates of synthetic $L_5$ Trojans accessible to Lucy during a pre-Patroclus encounter. These Trojans cluster on the sky $\sim$6 years prior to the encounter window. (\textbf{Bottom}) Time evolution of the search area for pre-Patroclus encounterable $L_5$ Jupiter Trojans. The sinusoidal-like lines represent the observable areas from Cerro Pachón, while the smooth transparent lines are the areas calculated in the barycentric frame. From bottom to top, the solid lines indicate, respectively, the 10$\%$, 50$\%$, and 80$\%$ percentile areas. The vertical green shaded band indicates the $\!\sim\!150\,\mathrm{d}$ search window. The lower right plot provides a zoom-in view centered on the 2026-2027 clustering window, which is marked in the left bottom plot with a box. The right-hand axis of the zoomed-in plot indicates the mean solar elongation of the synthetic Trojans. The mean solar elongation is shown as a gray dashed line, and the full range between minimum and maximum elongation is shaded in gray.}\label{fig:area}
\end{figure*}

For a pre-Patroclus encounter, after defining the state vectors of synthetic $L_5$ Trojans encounterable by Lucy with a $\Delta v =50 \, \mathrm{m/s}$ maneuver, we propagate their trajectories from Lucy’s launch date to the time of the Patroclus encounter. Their ecliptic coordinates, as observed from the Solar System barycenter, are shown in the top panel of Figure \ref{fig:area}. The synthetic Jupiter Trojans cluster in the sky not only once per orbital period ($\sim\!12$ years) but also halfway through their orbit ($\sim\!6$ years). This clustering arises because, during the encounter window, Lucy travels close to Jupiter’s orbital plane. Therefore, about half an orbit before the encounter window, these encounterable objects as they cross the opposite orbital node cluster together regardless of whether they are ascending or descending in their orbit. We refer to this period of enhanced clustering as the clustering window.

To characterize the search area, we quantify the projected spatial distribution of the encounterable objects using a Gaussian kernel density estimator (KDE). Rather than modeling just ecliptic latitude and longitude in 2D, we assign all coordinate sets a unit radius and perform 3D density modeling of the resulting Cartesian coordinates, constrained to the unit sphere. This approach mitigates projection effects and complications associated with cyclic coordinates. We evaluate the KDE on a uniform all-sky grid defined using the HEALPix tessellation \citep{Gorski:2005APJ} with pixels of $0.37\,\mathrm{deg}^2$ each (nside=96). The total search area is then defined by the number of pixels above a given KDE percentile threshold.

The time evolution of the search area for the 10$\%$, 50$\%$ and 80$\%$ percentiles of the distribution of synthetic pre-Patroclus Trojans encounterable with a $\Delta v=50\,\mathrm{m/s}$ is shown in the bottom panels of Figure \ref{fig:area}. A zoomed-in view around the clustering window is shown for the 10th percentile in the lower right panel. The topocentric sky distribution exhibits a sinusoidal pattern with a 1-year period, driven by geometric effects as the projected area expands when Earth moves closer to the $L_5$ cloud ($\sim\! r^{-2}$). The clustering window spans $\sim$1.5 years but it is partially unobservable because the $L_5$ cloud is behind the Sun during its extrema. The most favorable period for search occurs during the $\sim$5 month interval at the end of 2026, when the Trojans are near opposition, which we will refer to as the search window (green band in Figure~\ref{fig:area}).

A key challenge during this clustering window is that the projected search area does not shrink as tightly as it does during the actual encounter window (see the right minimum on the bottom left plot of Figure \ref{fig:area}). For example, during the clustering window, the 80\% percentile region has a minimum area of $\sim$900\,deg$^2$, while the 50\% percentile region reaches a minimum of $\sim$300\,deg$^2$. These areas are too large to justify a dedicated targeted search effort.

Instead, the most feasible approach is to focus on the $\sim$50\,deg$^2$ region corresponding to the 10\% percentile (bottom right panel of Figure~\ref{fig:area}). This strategy implies a 1-in-10 probability of identifying an additional flyby candidate within the accessible volume for a given combination of size limit $H_r$ and budget $\Delta v$ in Figure~\ref{fig:NE}. During the search window, this 10\% area varies between $\sim$30 and $\sim$50\,deg$^2$. Its rate of change is $\sim$0.3\,deg$^2$/day at the extrema and slows to $\sim$0.1\,deg$^2$/day in the central part.

We also found that the area covered by encounterable objects does not vary significantly across the range of $\Delta v$ values explored in this study. Additionally, we tested whether shifting the start of the encounter window (e.g., to when Lucy reaches the $1\,\sigma_r$ region of the radial distribution) would affect the area distribution. This adjustment led only to a shift in the timing of the clustering window, with no significant change to the overall area. These results suggest that the area distribution of encounterable objects is primarily set by the orbital element distribution of the $L_5$ Jupiter Trojan population, rather than by the specific value of $\Delta v$ or the timing of the encounter window.

The clustering properties of post-Patroclus encounterable Trojans exhibit similar characteristics to the pre-Patroclus case. The minimum clustering area (10th percentile) is $\sim$30 deg$^2$, slightly smaller than the pre-Patroclus minimum due to the relative orientation of the accessible volume as seen from Earth. However, this minimum occurs during early 2027, when the $L_5$ cloud is not observable. As a result, the optimal time to search for post-Patroclus encounterable Trojans coincides with that of the pre-Patroclus search: near opposition in late 2026. During this time, the clustering area is slightly larger, at $\sim$55-60 deg$^2$.

\begin{figure}[h]
    \centering
    \includegraphics[width=1\linewidth]{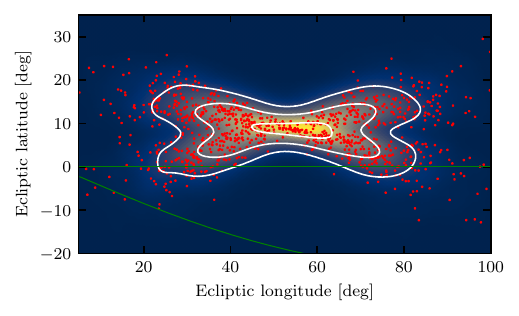}
    \caption{Projected sky distribution of encounterable pre-Patroclus $L_5$ Jupiter Trojans in ecliptic coordinates as observed from Cerro Pachón at the middle point of the search window. The white solid lines indicate the 10th, 50th, and 80th percentile regions and the red dots represent 1/10 of the synthetic Jupiter Trojans used to calculate them. The green solid lines are the ecliptic and equatorial planes.}
    \label{fig:searchwindow}
\end{figure}

Figure~\ref{fig:searchwindow} shows the sky distribution of pre-Patroclus encounterable $L_5$ Jupiter Trojans at the midpoint of the search window. The 10th percentile region corresponds to the portion of the sky where Trojans are crossing the orbital node opposite to the one associated with the encounter window. Notably, a significant fraction of the sky distribution of post-Patroclus encounterable Trojans overlaps with that of pre-Patroclus Trojans during opposition. This overlap arises from the higher local density of $L_5$ Trojans near the end of the pre-Patroclus encounter window and the beginning of the post-Patroclus window, which are separated by only $80$ days. As a result, a single observational campaign could support target identification for both scenarios.

\subsection{Survey duration} \label{subsec:survey}

There is a $\sim\!10\%$ probability of discovering an additional pre-Patroclus $L_5$ flyby target for Lucy with a moderate maneuver ($\Delta v \simeq 35-50 \, \mathrm{m/s}$), provided an area of $\!\sim\! 50\,deg^2$ is surveyed at any point within the $\sim\!5$-month search window. The telescope time required depends primarily on the target's size, which sets the necessary magnitude depth. For instance, assuming a 5$\%$ albedo, during the search window, a 1 $\mathrm{km}$ encounterable Trojan would have an apparent $r$-band magnitude of $m_r \!\sim\! 26$, whereas a 600 $\mathrm{m}$ object would reach $m_r \!\sim\! 27$. Thus, the ideal facility must combine high sensitivity (i.e., large aperture) with a wide FOV to minimize the number of pointings. At present, suitable facilities include the 4 m Victor M. Blanco telescope equipped with the $3.8\,\mathrm{deg^2}$ Dark Energy Camera (DECam, $2.2^\circ$ wide FOV) \citep{Flaugher:2015AJ}, and the 8.2 $\mathrm{m}$ Subaru telescope equipped with the $1.8\,\mathrm{deg^2}$ Hyper Suprime Cam (HSC, $1.5^\circ$ diameter FOV) \citep{Miyazaki:2018PASJ}. 

We also consider the Vera C. Rubin Observatory, expected to begin science operations by the end of 2025, which features the Simonyi Survey Telescope with a $6.5\,\mathrm{m}$ effective diameter primary mirror, and the $9.6\,\mathrm{deg^2}$ LSST Camera \citep{Ivezic:2019ApJ}. The LSST survey is projected to discover $\sim$100,000 Jupiter Trojans over its 10-year baseline \citep{Juric:2023AAS, Kurlander:2025AJ}, with a single-exposure depth of $m_r\sim24.5$, sufficient to detect Trojans down to $\sim1-2.5$ km. While most of these detections will likely occur in the early years of Rubin operations \citep{Kurlander:2025AJ}, the complete inventory is unlikely to be available by the time of the search window identified in Section \ref{subsec:whenwhere}. More importantly, as discussed in Section \ref{subsec:integral}, the Trojans most likely to be within the Lucy's accessible volume are $D\lesssim700$ m, beyond the reach of the Wide-Fast-Deep footprint. Therefore, we also evaluate how much dedicated time would be required for Rubin to conduct a targeted search.

\begin{figure}[h]
    \centering
    \includegraphics[width=1\linewidth]{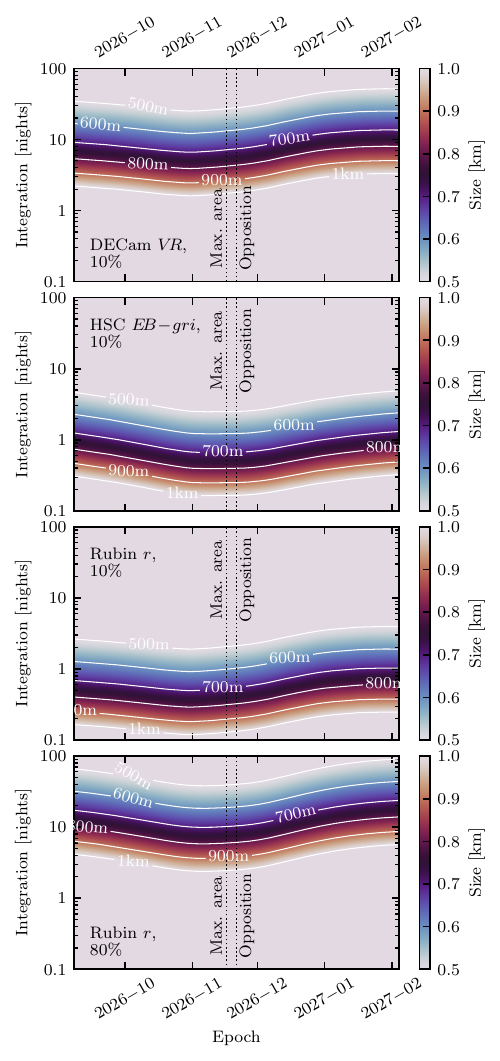}    
    \caption{Total integration time required to cover a given percentile of the sky distribution of pre-Patroclus $L_5$ encounterable Jupiter Trojans, down to a specified size (assuming $p$=0.05), and as a function of epoch within the search window. The top panel present results for DECam covering the 10th percentile region, while the second panel shows the same for Subaru/HST. The third and fourth panels correspond to the LSST Camera, considering both the 10th and 80th percentile regions. The colormap is truncated between $1 \, \mathrm{km}$ and $500 \, \mathrm{m}$. Vertical dotted lines indicate the epoch of maximum topocentric search area and the time of opposition.}
    \label{fig:surveynights}
\end{figure}

In this analysis, we estimate a lower limit on the survey duration required to detect an additional Lucy $L_5$ flyby target. This is an order of magnitude estimate, not a detailed survey plan. It assumes 5-$\sigma$ detections\footnote{Equation (6) in \cite{Ivezic:2019ApJ} for Rubin.}, that the HSC $EB-gri$ filter reaches $\sim$0.5 mag deeper than the $r2$ filter \citep{Fraser:2024PSJ}, and that the DECam $VR$ filter reaches $\!\sim\!0.1$ mag deeper than the $r$ filter, under dark and standard atmospheric conditions at each site. Figure \ref{fig:surveynights} illustrates the total integration time required to cover the sky distribution of pre-Patroclus encounterable Trojans down to a given size at a specific epoch. These integration times refer not to a single exposure, but to the cumulative exposure time needed to cover the full 10th percentile of the search area (and for Rubin, also the 80th percentile). Consequently, the required survey time reflects both the changing apparent brightness (due to evolving phase angles) and the time-dependent area of the search region, along with night length constraints defined between astronomical twilight and dawn at each site.

Our results indicate that the most favorable conditions occur near the midpoint of the search window. Even though the projected search area peaks at that time, the low phase angles yield significantly brighter apparent magnitudes. In practice, the optimal search conditions occur during the $\!\sim\!1$-month interval when the $10\%$ search area starts to decrease and the targets are near opposition.

Assuming the integration times in Figure \ref{fig:surveynights} serve as a proxy for the total survey duration, the Subaru/HSC system is, as expected, significantly more efficient than DECam for this search. Specifically, surveying the 10$\%$ region down to $700\, \mathrm{m}$ Trojan sizes near opposition with HSC would require less than one night of observing time, and $\!\sim\!1$ night at other phases (assuming a 5$\%$ albedo). By contrast, achieving the same detection with DECam would require $\!\sim\!10$ nights. For smaller targets, Subaru could cover the region down to $600\,\mathrm{m}$ objects in $\!\sim\!1-2$ nights, while reaching $500\,m$ targets would require a more extensive campaign of $\!\sim\!2-5$ nights. If Rubin were available for this search, less than one night would be sufficient to survey the 10th percentile region and detect encounterable targets as small as $600\,\mathrm{m}$ in a single night. Reaching $500\,\mathrm{m}$ targets would require 2 to 3 nights. Moreover, Rubin could scan the entire 80th percentile region down to a depth sufficient to detect $700\,\mathrm{m}$ objects in $\!\sim\!10$ nights. Comparable values were obtained when considering only encounterable Trojans in the post-Patroclus scenario.

\section{Discussion} \label{sec:discussion}

\subsection{Science return from a sub-km $L_5$ Trojan}

Our analysis indicates that there is a $\sim$7-10$\%$ probability of identifying an additional sub-kilometer (500$-$700 m) $L_5$ Jupiter Trojan flyby target for Lucy, achievable with a reasonable investment of resources. The largest primary Lucy flyby target is the binary Patroclus ($106\,\mathrm{km}$) and Menoetius ($98\,\mathrm{km}$) \citep{Mueller:2010Icar}, while the smallest is Polymele ($\sim\!20\,\mathrm{km}$). Including a sub-kilometer Trojan target would significantly enhance the mission by expanding the size range of primary Trojan targets. The striking and unexpected results from the Lucy encounter with the $\!\sim\!700$ m diameter MBA Dinkinesh validate the potential of such an encounter \citep{Levison:2024Natur}.

A Trojan in the sub-kilometer range would likely be a collisional fragment. In this regard, it may share characteristics with the $L_4$ target Eurybates, which, despite its larger size, with a surface-equivalent spherical diameter of $69.3 \pm 1.4 \, \mathrm{km}$ \citep{Mottola:2023PSJ}, is the largest member of the most prominent collisional family of Jupiter Trojans \citep{Broz:2011MNRAS, Vokrouhlicky:2024AJ}. Notably, Eurybates host a kilometer-scale satellite, Queta ($1.2\pm0.4\,\mathrm{km}$) \citep{Noll:2020PSJ, Brown:2021PSJ}, and Polymele may also harbor a satellite with an estimated diameter of $\!\sim\!5-6\,\mathrm{km}$, inferred from stellar occultations \citep{Buie:2022DPS}. While both Queta and Polymele's satellite may be observed during Lucy's encounters, an in situ study of an isolated sub-$\mathrm{km}$ Jupiter Trojan from the background population would provide unique and independent insights. Such an observation would enable a more comprehensive comparison of the collisional environment and evolutionary histories of the $L_4$ and $L_5$ populations.

Lucy could offer a opportunity to directly explore the $L_4/L_5$ asymmetry if an additional sub-kilometer Trojan is included in the $L_5$ flyby schedule. The largest Trojans (up to $H_V \!\sim\!9.5$), corresponding to $\gtrsim 80\,\mathrm{km}$ objects, are symmetric within statistical uncertainties \citep{Vokrouhlicky:2024AJ}. This means that if Patroclus remains Lucy's only $L_5$ target, the mission would not be positioned to directly address the asymmetry problem, as the only way to constrain the properties of small $L_5$ Trojans would be indirectly via cratering records. As noted by \cite{Vokrouhlicky:2024AJ}, the observed asymmetry in magnitude-limited samples could be explained by a fractional difference in mean albedo which is not supported by WISE/NEOWISE constraints \citep{Grav:2011ApJ, Grav:2012ApJ}. A plausible explanation is that the asymmetry arises from differences in shape distributions between the two clouds. This is supported by observations from ATLAS, which suggest that on average $L_4$ Trojans are more elongated than $L_5$ Trojans \citep{McNeill:2021PSJ}. Although a single small Trojan from the $L_5$ cloud cannot represent the entire population statistically, in situ spacecraft observations can nevertheless yield a wealth of information about its shape, composition, and surface properties, and thus offer valuable clues to the broader cloud characteristics. This means that a direct observation of a sub-kilometer Jupiter Trojan could strengthen Lucy's ability to investigate the $L_4/L_5$ asymmetry by enabling a direct comparison of the properties of small Trojans from both clouds.

\subsection{Sensitivity to the faint-end of the $H$-distribution}\label{subsec:hsensi}

The most uncertain component of our $L_5$ cloud modeling is the shape of the cumulative absolute magnitude distribution. For simplicity, we have assumed a broken power law with two breaks (Section \ref{subsec:Hdistr}). However, recent work has introduced a cubic-spline representation that better captures local variations in the slope \citep{Vokrouhlicky:2024AJ}. While this spline approach may be more appropriate for describing the large size regime where the slope variations are larger, our simplified power law is adequate for the small-size end, which dominates the probability of identifying an additional $L_5$ flyby target (Section \ref{subsec:integral}). As such, our predicted number of encounterable Trojans is most sensitive to the location of the faint-end break and the rollover slope. 

Recent studies start to hint that the $\sim$0.3 faint-end slope observed for small Jupiter Trojans down to $D\!\sim\!2\,\mathrm{km}$ \citep{Uehata:2022AJ, Yoshida:2017AJ} (which is the nominal value in our modeling), may continue down to the $D\!\sim\!1$ km detection limit of recent surveys \citep{Chen:2023DPS}. These HSC-based studies find similar faint-end slopes for both the $L_4$ and $L_5$ clouds down to $D\!\sim\!2$ km. In contrast, \cite{Vokrouhlicky:2024AJ} report a potentially shallower slope for the $L_5$ population, although they caution that this measurement may be unreliable due to their small number of detections in that range.

\begin{figure}
    \centering
    \includegraphics[width= \linewidth]{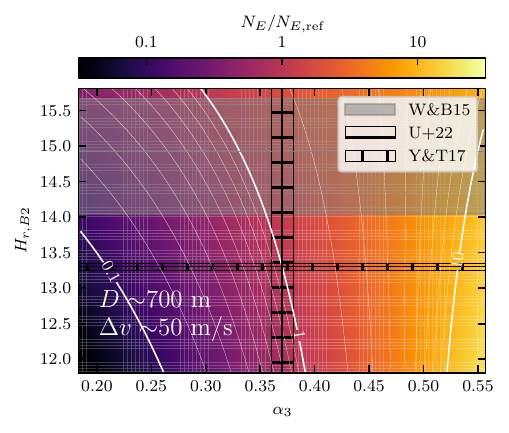}
    \caption{
    Expected number of encounterable $L_5$ Jupiter Trojans under different assumptions for the faint-end break magnitude $H_{r,B2}$ and the corresponding power-law slope $\alpha_3$. The values are normalized to a reference case assuming $D=700$ m (p=5$\%$) and $\Delta v=50$ m/s, which yields $N_E \!\sim\! 1$ during a pre-Patroclus scenario. Hashed vertical and horizontal bars indicate uncertainty ranges from our nominal values, i.e. from \cite{Uehata:2022AJ} and \cite{Yoshida:2017AJ}, respectively. The horizontal gray line and shaded band represent the break magnitude reported by \cite{Wong:2015AJ}.
    }
    \label{fig:variations}
\end{figure}

Our prediction could be most affected by the location of the power-law break, as it has not yet been directly constrained for the $L_5$ cloud. For the $L_4$, \cite{Wong:2015AJ} report a break at $H_{B2}=14.93_{-0.88}^{+0.73}$, while \cite{Yoshida:2017AJ} find a break at $H_{B2}=13.56_{-0.06}^{+0.04}$. We adopt the latter as our nominal value for $L_5$ (Section \ref{subsec:Hdistr}), since the only published deep survey of the $L_5$ population \citep{Uehata:2022AJ} does not constrain the break location. \cite{Vokrouhlicky:2024AJ} propose that the break may be smoother than allowed by a broken power-law model, which could explain the discrepancy between the reported measurements. Although their analysis suggests a transition to a shallower slope at fainter magnitudes for the $L_5$ cloud, they also note this regime lies near the detection limit of their data, adding considerable uncertainty.

To explore the sensitivity of our predictions to the shape of the magnitude distribution, we evaluate how changes in the faint-end break and slope affect the number of encounterable Trojans. This is illustrated in Figure \ref{fig:variations}. We use as a reference case the expected number of Trojans accessible in the pre-Patroclus window with $\Delta v = 50$ m/s and $D = 700$ m (assuming 5$\%$ albedo), which yields $N_E \!\sim\! 1$ (see Section \ref{subsec:integral}). We vary $\alpha_3$ between 0.5 and 1.5 times our nominal value, and shift the break location $H_{B2}$ by $-1.5$ and $+2.5$ magnitudes relative to our nominal value. As shown in Figure \ref{fig:variations}, if the break occurs at fainter magnitudes, as suggested by \cite{Wong:2015AJ} and \cite{Vokrouhlicky:2024AJ}, the expected number of accessible Trojans could double. Although variations in $\alpha_3$ can also have a substantial impact, the existing studies largely agree on its approximate value, suggesting limited room for drastic departures.

Our predictions may also be affected by the magnitude distribution at sizes below $D < 1$ km, a regime not yet constrained observationally. It has been proposed that non-gravitational forces could lead to a shallower slope for objects between 0.1–1 $\mathrm{km}$ in size. This is due to the Yarkovsky force driven by diurnal heating and asymmetric thermal reradiation which could induce orbital destabilization of small Jupiter Trojans \citep{Hellmich:2019A&A}. This possibility underscores the need to extend the characterization of the absolute magnitude distribution of Jupiter Trojans to sub kilometric scales, both to confirm the location and slope of collisional breaks \citep{Marschall2022AJ}, and to test the depletion of small objects due to non-gravitational effects.

The interpretation of our absolute magnitude distribution depends on the assumed albedo.  As discussed in Section~\ref{subsec:Hdistr}, it is unclear if the reported increase in albedo among smaller Trojans is a bias or a systematic effect, but if the latter is true, adopting $p=0.05$ for all sizes would be an oversimplification.  Although this does not alter the $H_r$ distribution itself, it does impact the diameter inferred from a given $H_r$: higher albedo implies smaller diameter.  For instance, $H_r=19.5$ (which at $\Delta v\sim\ 50 \, \mathrm{m/s}$ yields $N_E\sim1$ in Figure~\ref{fig:NE}) corresponds to $D\sim700\ \mathrm{m}$ for $p=0.05$, $D\sim600\ \mathrm{m}$ for $p=0.07$, and $D\sim500\ \mathrm{m}$ for $p=0.09$.  Similarly, in Figure~\ref{fig:surveynights}, the survey durations plotted for 1 km–500 m objects (assuming $p=0.05$) would instead apply to $\sim850\ \mathrm{m}$ – $400\ \mathrm{m}$ objects if $p=0.07$.  These adjustments do not change our conclusions about flyby target probability, but they imply that accessible Trojans could be even smaller than initially inferred.

\subsection{Survey strategy and observational considerations}

Several practical considerations for future planning are worth noting despite designing the actual survey is beyond the scope of this work. A key caveat of our lower-limit estimates (Figure \ref{fig:surveynights}) is that they represent only the total integration time under standard conditions and do not account for factors such as image quality or readout time overheads. For instance, during the Arrokoth search, observations conducted at high air masses ($>1.6$) or under non-photometric conditions were inadequate to achieve the search goals \citep{Buie:2024PSJ}. 

In standard sidereal tracking, moving objects will trail in the image with a length that depends on the integration time. To avoid SNR degradation, the exposure time must be limited so that the trailing does not exceed the PSF size ($\propto  t^{-1/2}$) \citep{Heinze:2015AJ}. The maximum allowable exposure time is then set by the object's sky motion, which varies with phase angle. For example, during opposition within the search window,  encounterable Jupiter Trojans exhibit projected sky motions as high as $\!\sim\!20 \mathrm{''/h}$. At larger phase angles, the projected sky motion decreases as the radial velocity component increases. HSC has a total overhead of $\!\sim\!40\,\mathrm{s}$ per frame, while DECam's is $\!\sim\!20\,\mathrm{s}$. These overheads could inflate the total survey time reported in Figure \ref{fig:surveynights} by a factor up to $\!\sim\!1.5$. A possible approach, analogous to \cite{Buie:2024PSJ} Hubble search, is to set the telescope to track at rates representative of the motion of encounterable Trojans, enabling longer exposures and reducing trailing losses. However, non-sidereal tracking at the high sky rates of Trojans introduces additional complexities in tiling the search area, maintaining accurate telescope guiding, and processing the data in later stages.

Under optimal conditions with sidereal tracking, the Subaru/HSC facility could detect $D\!\sim\!1\,\mathrm{km}$ Trojans in single exposures (e.g., \citealt{Chang:2021PSJ}). However, for targets in the range of interest ($700 > D \geq 500 \,\mathrm{m}$), the objects would likely fall below the single exposure detection threshold. In these cases, the shift-and-stack technique is required, where multiple short exposures are aligned along the object's motion and co-added to recover the $\propto t^{1/2}$ SNR proportionality. The shift-and-stack technique has been widely applied, for example, in deep wide-area surveys of KBOs (e.g., \citealt{Napier:2024PSJ}), and early results from its application to Jupiter Trojan science are encouraging (e.g. \citealt{Napier:2023thesis}, \citealt{SalazarManzano:2024}).  

An effective search strategy would ideally include a parallel astrometric follow-up effort, either using the same search telescopes or separate facilities, with the latter not requiring large FOV instruments. This means data processing and source detection would need to be performed on short timescales. A modern shift‐and‐stack survey differs from traditional tracklet searches because preliminary orbit estimates are built into the detection itself. By spreading the required exposures for each field over multiple nights, the method could yield Trojan candidates with an inferred $\Delta v$ for a Lucy intercept.  Future work will establish the optimal $\Delta v$ cutoff for follow‐up, but even so, the telescope time needed to confirm `below‐threshold' candidates should remain only a small fraction of the discovery time shown in Figure~\ref{fig:surveynights}.  Given the six-year lead time before the encounter window, and the four-year lead time before EGA3, additional observations around the late-2028 opposition would likely be required to achieve the orbital precision necessary for confirming a viable Lucy intercept and to determine the characteristics of the EGA3 maneuver.

If a flyby candidate is identified and integrated into Lucy's mission plan, it would require a multi-instrument, multi-wavelength follow-up characterization campaign. Pre-encounter characterization is key to optimizing encounter geometry and science return (e.g., \citealt{Wong:2024PSJ, Noll:2020PSJ, Mottola:2023PSJ, Mottola:2020PSJ, Buie:2021PSJ, Buie:2020AJ, Buie:2018AJ}), particularly for a sub-kilometer Trojan. The ongoing drift of the $L_5$ cloud towards lower galactic latitudes opens up possibilities for stellar occultation campaigns, which can deliver high-precision constraints on size and shape \citep{Buie:2020AJ}, as well as detect satellites or binary companions. These campaigns do, however, depend on accurate orbits \citep{Porter:2018AJ, Porter:2022PSJ}, which could be difficult to establish for very small objects. For a sub-kilometer object, an occultation campaign would be challenging due to the small station spacing required in the cross-track direction. Ultimately, space-based facilities such as the Hubble Space Telescope and JWST may be required for refining the astrometric orbit and inferring diameter, shape, albedo, colors, thermal properties, spin orientation, and rotation period. Since the potential flyby would occur in 2032-2033, next-generation observatories such as the Extremely Large Telescope (ELT) and the Nancy Grace Roman Telescope could also play a role in pre-encounter characterization.

\subsection{Future opportunities}

Modification of Lucy's primary mission schedule is challenging by construction since its architecture was optimized for the very specific sequence encounter characteristics \citep{Olkin:2024SSRv}. Even if a search discovers new targets that ultimately do not fit Lucy's primary timeline, those objects still hold value for remote spacecraft observations. From Earth, Jupiter Trojans reach maximum phase angles of $\!\sim\!15^\circ$ (e.g. \citealt{Schaefer:2010Icar, Schemel:2021PSJ}), whereas a spacecraft vantage point can achieve large phase angles, which is useful to constrain surface properties \citep{Porter:2016ApJ, Verbiscer:2019AJ, Verbiscer:2022PSJ}.

The search for Arrokoth \citep{Spencer:2003EM&P, Buie:2024PSJ} demonstrated that systematically finding unknown flyby targets is indeed feasible. In this work, we have shown how the dynamics of resonant populations in the Solar System create favorable conditions for discovering additional flyby options, and we have outlined a methodology for evaluating their viability. The ``nodal clustering'' these resonant populations exhibit twice per orbit could be used to find new Jupiter Trojan targets in both the $L_5$ and $L_4$ clouds for possible extensions of the Lucy mission. Notably, the clustering that occurs one full orbital period before the encounter window is tighter than the clustering at half an orbit, highlighting the importance of conducting this type of analysis well in advance. This approach can also be applied to other mean-motion resonances interior to Lucy's orbit, such as Hildas (3:2), which are likewise believed to contain primitive remnants from the outer Solar System \citep{Vokrouhlicky:2025AJ, Chang:2022ApJS}, and Thules (4:3), located at the outermost edge of the asteroid belt. This also suggests a broader strategy: with sufficient lead time, resonant populations throughout the Solar System could be systematically searched to identify encounterable targets tailored to the trajectories of current and future flyby missions.

\section{Summary / Conclusions} \label{sec:conclusions}

We have investigated the feasibility of discovering and flying by an additional, yet unknown, $L_5$ Jupiter Trojan with NASA's Lucy spacecraft. The main components of this work and findings are summarized as follows:

\begin{enumerate}

\item  We model the $L_5$ cloud using the known background Trojans brighter than the current completeness limit as representative of the entire population. The absolute magnitude distribution is described by a power law with two breaks, scaled to match the observed distribution (Figure \ref{fig:Hdistr}). The spatial distribution is constructed in the Jupiter co-rotating frame, using a lognormal distribution for the longitude and Gaussian distributions for both the latitude and heliocentric distance (Figure \ref{fig:spatialdistr}). This three-dimensional approach improves upon previous two-dimensional Gaussian approximations by capturing both the asymmetry in longitude and the radial oscillations due to perihelion-aphelion motion. Through appropriate coordinate transformations, our model enables calculation of the number density of $L_5$ Jupiter Trojans brighter than a given absolute magnitude in a given position in space. 

\item We consider two encounter opportunities, pre- and post-Patroclus, defined by the times Lucy enters and exits the spatial extent of the $L_5$ cloud, without affecting the scheduled time or position of the Patroclus flyby (Figures \ref{fig:windows} and \ref{fig:Devol}). The pre-Patroclus scenario spans $\!\sim\!250$ days and involves a ``double-cone'' accessible volume enabled by a cost-free maneuver at EGA3. The post-Patroclus scenario spans a longer $\sim650$-day window with a single-cone geometry extending from shortly after the Patroclus encounter.  

\item We compute the expected number of encounterable $L_5$ Trojans for a given $\Delta v$ budget using a semi-analytical method (Figure \ref{fig:NE}). Our results indicate that no Trojans down to $100\,\mathrm{m}$ are expected in the volume accessible along Lucy's nominal trajectory. However, moderate maneuvers ($\Delta v
\!\sim\! 35-50$ m/s) are expected to yield at least one encounterable sub-kilometer Trojan ($D\!\sim\!500-700$ m) in the pre-Patroclus window. Due to the smaller spatial densities after the Patroclus encounter, the probability of encountering a $D\!\sim\!700-500$ m Trojan in the post-Patroclus window is $\sim40-70\%$.

\item Simulation of synthetic Trojans reveal a ``nodal clustering'' effect occurring twice per orbit, which creates favorable conditions for a targeted search $\!\sim\! 6$ years prior to the encounter window (Figure \ref{fig:area}). For both the pre- and post-Patroclus cases, our analysis suggests that the optimal search period is a 5-month window toward the end of 2026. Although the total area remains sizable, the most efficient strategy is to target the $10\%$ highest sky density region, which spans $\!\sim\!50\,\mathrm{deg}^2$. Notably, the size of the search area is primarily determined by the Keplerian elements of the Trojan population, rather than by the $\Delta v$ budget or the precise timing of the encounter window.

\item A single observing campaign could support the identification of targets for both the pre- and post-Patroclus scenarios. Scanning the high-density $10\%$ sky distribution to sufficient depth to detect pre-Patroclus Trojans between $700\, \mathrm{m}$ and $500\,\mathrm{m}$ could be accomplished in fewer than 5 nights using the Hyper Supreme Cam on the Subaru Telescope with the $EB-gri$ filter (Figure \ref{fig:surveynights}). The same search down to $600\,\mathrm{m}$ objects could be completed in a single night with Rubin. Achieving the required depths will likely require the use of the shift-and-stack technique to co-add multiple short exposures and reach the faint magnitude limits of the targets.

\item The discovery of a sub-kilometer Jupiter Trojan, which we estimate has a $\sim\!7\!-\!10\%$ probability when accounting for both spatial accessibility and observational constraints, would significantly enhance the scientific return of the Lucy mission. It would extend the size range of primary Trojan flyby targets and provide a rare opportunity to compare the physical properties of  small $L_4$ and $L_5$ Trojans, shedding light on their collisional environments and the origins of the observed asymmetry between the two clouds.

\item Even if a flyby target is not incorporated into the primary mission, a dedicated search may yield candidates suitable for remote observations at high phase angles, a geometry inaccessible from Earth. Moreover, the methodology presented in this work is broadly applicable to other searches for flyby targets in the Jupiter Trojan population (both $L_5$ and $L_4$), Hildas, or other resonant populations of interest to current and future planetary missions. 

\end{enumerate}

Taken together, these results highlight that a systematic survey targeting nodal clustering in the $L_5$ cloud offers a meaningful probability of discovering an additional Lucy flyby target. A sub-kilometer object in an accessible orbit would expand Lucy’s scope, revealing important insights into the diversity of the Trojan population as primitive remnants from the outer Solar System.


\section{Acknowledgements} \label{sec:acknow}
We thank our anonymous reviewers for their constructive feedback, which helped improve the quality of this work. This material is based upon work supported by the National Science Foundation under grant No.\ AST-2406527. This research was supported in part through computational resources and services provided by Advanced Research Computing at the University of Michigan, Ann Arbor. Tessa Frincke acknowledges support from NSF grant number AST-2303553. Luis Salazar Manzano gratefully acknowledges useful discussions and insights from Aster Taylor, Gabriel Patron, Pedro Bernardinelli, and Thomas Ruch. We thank Hal Levison for insightful conversations that inspired this study. 


%

\vspace{5mm}


\software{spacerocks \citep{Napier:2020spacerocks}, astropy \citep{Astropy:2013A&A, Astropy:2018AJ}, scipy \citep{SciPy:2020}, spiceypy \citep{spiceypy:2020}, healpix \citep{Gorski:2005APJ}.}



\appendix

\begin{figure*}
    \centering
    \begin{minipage}{\textwidth}
        \centering
        \includegraphics[width=\linewidth]{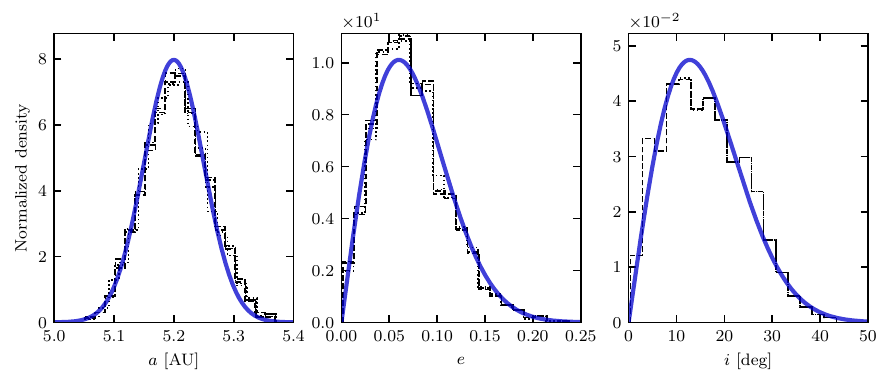}   
    \end{minipage}
    \caption{Orbital distribution model of the $L_5$ Jupiter Trojans. The three panels show the distributions of semi-major axis ($a$), eccentricity ($e$), and inclination ($i$), respectively. Histograms correspond to background Trojans from the MPC catalog that are brighter than the completeness limit, evaluated at four different epochs. The blue curves represent fitted models, with best-fit parameters: $\mu_a = 5.2$ AU, $\sigma_a = 0.05$ AU, $\sigma_e = 0.06$, and $\sigma_i = 12.78^\circ$.}    
    \label{fig:orbitaldistr}
\end{figure*}

\section{Cylindrical Coordinate System Along a Spacecraft Path} \label{app:cylcoords}

To parametrize the space around a spacecraft following a known trajectory, a practical method is to define cylindrical slices perpendicular to the instantaneous direction of motion.  A point in this coordinate system is described by the parameters $t$, $\theta$, and $\rho$, where:

\begin{itemize}
    \item $t$ specifies the spacecraft’s position $\mathbf{r}(t)$ and velocity $\mathbf{v}(t)$ at a given time,
    \item $\theta$ is the azimuthal angle within the perpendicular plane, with $0 \leq \theta < 2\pi$, and
    \item $\rho$ is the radial distance from the spacecraft to a point in the plane.
\end{itemize}

\begin{figure}
    \centering
    \includegraphics[width=1\linewidth]{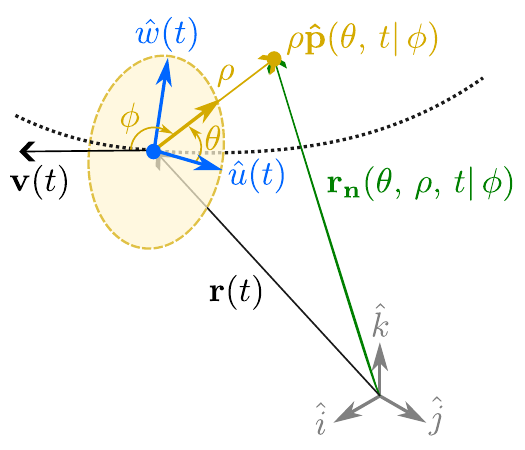}
    \caption{Schematic diagram of the local cylindrical coordinate system along a spacecraft's trajectory. The vectors $\mathbf{\hat{u}}(t)$ and $\mathbf{\hat{w}}(t)$ define the plane orthogonal to the velocity vector $\mathbf{v}(t)$ when $\phi=90^\circ$.}
    \label{fig:cylindricalcoords}
\end{figure}

A spherical generalization introduces the polar angle $\phi$, which describes the angle between the velocity vector and the radial direction. This generalization is particularly useful when considering $\Delta\mathbf{v}$ maneuvers. A schematic of the coordinate system is shown in Figure~\ref{fig:cylindricalcoords}.

The first orthonormal unit vector is:

\begin{equation}
    \mathbf{\hat{u}}(t) = \frac{\hat{\mathbf{k}} \times \mathbf{v}(t)}{\left\| \hat{\mathbf{k}} \times \mathbf{v}(t) \right\|}
    = \frac{\langle -v_y(t), v_x(t), 0 \rangle}{v_{xy}(t)},
\end{equation}

where the projection of the velocity vector in the x-y plane is $v_{xy}(t) = \sqrt{v_x^2(t) + v_y^2(t)}$. The second orthonormal unit vector is:


\begin{equation}
    \mathbf{\hat{w}}(t) \! = \! \frac{\mathbf{v}(t) \! \times \! \mathbf{\hat{u}}(t)}{\|\mathbf{v}(t) \! \times \! \mathbf{\hat{u}}(t)\|}
   \! = \! \frac{\langle 
    -v_x(t) v_z(t), \; -v_y(t) v_z(t), \; v_{xy}^2(t)
    \rangle}{\|\mathbf{v}(t)\| v_{xy}(t)}.
\end{equation}


Where the magnitude of the velocity is $\|\mathbf{v}(t)\| = \sqrt{v_x^2(t) + v_y^2(t) + v_z^2(t)}$. The direction of a point relative to the spacecraft, within the perpendicular plane ($\phi=90^\circ$), or more generally in the spherical case, is given by the following expressions:


\begin{align}
    &\mathbf{\hat{p}}(\theta, t) = \cos\theta \, \mathbf{\hat{u}}(t) + \sin\theta \, \mathbf{\hat{w}}(t), \\
    &\mathbf{\hat{p}}(\theta, t \mid \phi) = \cos\theta \sin\phi \, \mathbf{\hat{u}}(t) + \cos\phi \, \mathbf{\hat{v}}(t) + \sin\theta \sin\phi \, \mathbf{\hat{w}}(t).
\end{align}




Where the unit vector along the velocity vector and perpendicular to this plane is simply $\mathbf{\hat{v}}(t) = \mathbf{v}(t)/\|\mathbf{v}(t)\|$. The corresponding expressions for both cases are expanded below:


\begin{equation}
\begin{split}
\mathbf{\hat{p}}(&\theta,t) =  \frac{1}{v_{xy}(t)} \biggl\langle
  -v_y(t) \cos\theta 
  - \frac{v_x(t) v_z(t)}{\|\mathbf{v}(t)\|} \sin\theta, \\ &\quad v_x(t) \cos\theta 
  - \frac{v_y(t) v_z(t)}{\|\mathbf{v}(t)\|} \sin\theta, \quad \frac{v_{xy}^2(t)}{\|\mathbf{v}(t)\|} \sin\theta
\biggr\rangle.
\end{split}
\end{equation}

And: 

\begin{equation}
\begin{split}
\hat{\mathbf{p}}(&\theta, t \mid \phi) = 
\frac{1}{\|\mathbf{v}(t)\|} \\ \biggl\langle 
&\cos\phi \, v_x(t) 
- \frac{\sin\phi}{v_{xy}(t)} \left( 
    \cos\theta \, v_y(t) \, \|\mathbf{v}(t)\| 
    + \sin\theta \, v_x(t) v_z(t)
\right), \\
&\cos\phi \, v_y(t) 
+ \frac{\sin\phi}{v_{xy}(t)} \left( 
    \cos\theta \, v_x(t) \, \|\mathbf{v}(t)\| 
    - \sin\theta \, v_y(t) v_z(t)
\right), \\
&\cos\phi \, v_z(t) 
+ \sin\phi \, \sin\theta \, v_{xy}(t)
\biggr\rangle
\end{split}
\label{eq:punit}
\end{equation}





This defines the transformation from cylindrical coordinates $(\theta, \rho, t |\,\phi)$ to Cartesian coordinates $(x, y, z)$ as:

\begin{equation}
    \mathbf{r_n}(\rho,\theta,t|\,\phi) =\mathbf{r}(t) + \rho \, \mathbf{\hat{p}}(\rho,\theta,t|\,\phi) 
\label{eq:rn}
\end{equation}

This system is valid as long as the trajectory does not lie entirely along the $z$-axis, which is rarely the case in planetary missions assuming that $x$-$y$ defines the ecliptic plane.

\section{Accessible Volume Optimization} 
\label{app:optimization}

In this section, we compute the volume accessible to Lucy for potential flybys during the pre- and post-Patroclus encounter windows (Section~\ref{subsec:windows}). We adopt the cylindrical coordinate system from Appendix~\ref{app:cylcoords}, where the accessible volume is defined by the radial extent $\rho$ and azimuthal angle $\theta$ in the plane perpendicular to Lucy's velocity vector at each time $t$. Importantly, this volume will be parametrized by the $\Delta v$ cost. For the pre-Patroclus scenario, Lucy can receive an effective $\Delta v$ during EGA3 without fuel expenditure, but a return maneuver is still required to redirect Lucy back to Patroclus. In contrast, for the post-Patroclus case, the full maneuver cost comes from the single impulse used to open the cone of accessible space, since no return constraint exists.

To frame the pre-Patroclus case, we define: $t_A$ as the time of EGA3, $t_B$ as the return maneuver time, and $t_C$ as the Patroclus encounter. Lucy's position and velocity at $t_A$ define the initial state $\mathbf{r}_A$ and $\mathbf{v}_A$. The free parameters are the direction and magnitude of the initial impulse $\Delta\mathbf{v}_A$, the return time $t_B$, and the required return impulse $\Delta\mathbf{v}_B$. A complication is that although Lucy's state at $t_B$ is fixed by the first maneuver, $\Delta\mathbf{v}_B$ must be computed to satisfy the boundary condition of returning to Patroclus at the nominal time $t_C$ and location.

To build intuition, we first consider a simplified linear toy model with constant speed $v_z=(z_C-z_A)/(t_C-t_A)$ along the $\hat{z}$ axis. The initial and return maneuvers are assumed to be perpendicular to the direction of motion. In this case, $\Delta v_A$ and $\Delta v_B$ are linearly related: $\Delta v_A = \Delta v_B(t_C - t_B)/(t_C - t_A)$. Although this is a very simplified approximation of the problem, it provides three key insights that hold in the more general case: (1) the timing of the return maneuver is independent of its amplitude, (2) this time can be determined by maximizing the area within the encounter window, and (3) the amplitude of the opening maneuver follows a simple relation set by a fixed return maneuver amplitude.

In the real problem, Lucy’s motion is non-linear, lacks azimuthal symmetry, and the opening and return maneuvers are not restricted to being perpendicular to the direction of motion. To compute the pre-Patroclus accessible volume, we assume Lucy's motion is governed solely by the gravitational influence of Sun. In this two-body framework, the position and velocity of the spacecraft at any time $t$ are given by $\mathbf{r}(t)=f(t,\,t_0)\,\mathbf{r_0}+g(t,\,t_0)\,\mathbf{v_0}$ and $\mathbf{v}(t)=\dot{f}(t,\,t_0)\,\mathbf{r_0}+\dot{g}(t,\,t_0)\,\mathbf{v_0}$, respectively. Here, $f$ and $g$ are the Gauss ``$f$'' and ``$g$'' functions, and $\dot{f}$, $\dot{g}$, are their time derivatives. These functions are fully determined by solving the universal Kepler's equation, given the initial position $\mathbf{r_0}=\mathbf{r}(t_0)$ and velocity $\mathbf{v_0}=\mathbf{v}(t_0)$ at time $t_0$ \citep{Danby:1992}.

The nominal Lucy trajectory, defined by the initial conditions at EGA3 ($t_A$), in addition to the corresponding velocity vector along this orbit are given by:   

\begin{align}
    \mathbf{r}_1(t)=f_1(t,\,t_A)\,\mathbf{r}_{1,A}+g_1(t,\,t_A)\,\mathbf{v}_{1,A} \label{eq:r1}\\
    \mathbf{v}_1(t)=\dot{f}_1(t,\,t_A)\,\mathbf{r}_{1,A}+\dot{g}_1(t,\,t_A)\,\mathbf{v}_{1,A} \label{eq:v1}
\end{align}    

The trajectory and evolution of the velocity vector, assuming an effective maneuver is executed during the gravity assist, are given by:

\begin{align}
    \mathbf{r}_2(t|\theta_A,\phi_A,\Delta v_A)
        &= f_2(t,t_A)\,\mathbf{r}_{1,A} \notag \\
        &+ g_2(t,t_A)\,\mathbf{v}_{2,A}(\theta_A,\phi_A,\Delta v_A) 
        \label{eq:r2} \\
    \mathbf{v}_2(t|\theta_A,\phi_A,\Delta v_A)
        &= \dot{f}_2(t,t_A)\,\mathbf{r}_{1,A} \notag \\
        &+ \dot{g}_2(t,t_A)\,\mathbf{v}_{2,A}(\theta_A,\phi_A,\Delta v_A)
        \label{eq:v2}
\end{align}

The post-maneuver velocity vector resulting from the gravity assist can be expressed as:

\begin{equation}
    \mathbf{v}_{2,A}(\theta_A,\,\phi_A,\,\Delta v_A)=\mathbf{v}_{1,A}+\Delta v_A\,\mathbf{\hat{p}}(\theta_A,\,\phi_A,\,t_A)
\end{equation}

This formulation assumes an instantaneous maneuver. The unit vector $\mathbf{\hat{p}}$ is defined in Equation~\eqref{eq:punit}. This representation conveniently separates the maneuver’s amplitude from its direction, allowing the problem to be parameterized by the azimuthal angle $\theta$ within the cross-sectional plane while still accommodating maneuvers with varying polar angles $\phi$.

If a second maneuver is executed at a later time $t_B$, the resulting trajectory is given by:

\begin{equation}
\begin{split}
    \mathbf{r}_3(t|\theta_A, \phi_A,\Delta v_A, t_B)=&f_3(t,\,t_B)\,\mathbf{r}_2(t_B|\theta_A,\phi_A,\Delta v_A)\\+&g_3(t,\,t_B)\,\mathbf{v}_{3,B}
\end{split}
\label{eq:r3}
\end{equation}

The position of the modified trajectory at the return maneuver time, $\mathbf{r}_2(t_B|\theta_A,\phi_A,\Delta v_A)$, is given by Equation \eqref{eq:r2}. The impulse required to redirect the spacecraft back to Patroclus is:

\begin{equation}
    \Delta v_B(\theta_A,\phi_A,\Delta v_A, t_B)=\lVert \mathbf{v}_{3,B}-\mathbf{v}_{2}(t_B|\theta_A,\phi_A,\Delta v_A)\rVert
\label{eq:deltavb}
\end{equation}

Where the velocity from the initially perturbed trajectory at the time of the return maneuver $\mathbf{v}_{2}(t_B|\theta_A,\phi_A,\Delta v_A)$ is obtained with Equation \eqref{eq:v2}. 
 
Determining the magnitude ($\Delta v_B$) and direction required to return the spacecraft to Patroclus is a spacecraft targeting problem, that can be addressed by solving Lambert's problem. Lambert's problem is a boundary-value formulation in which two position vectors and the time of flight between them are known. The unique elliptical orbit that connects these points is found by solving Lambert's equation via the bisection algorithm \citep{Prussing:1993}. In our case, the position vector at the return maneuver time is $\mathbf{r}_3(t_B|\theta_A, \phi_A, \Delta v_A)=\mathbf{r}_2(t_B|\theta_A, \phi_A, \Delta v_A)$, and the target position and time for the Patroclus encounter are given by $\mathbf{r}_3(t_C)=\mathbf{r}_1(t_C)$, with the latter computed using Equation \eqref{eq:r1}. Solving Lambert's problem provides the required velocity at $t_B$, $\mathbf{v}_{3,B}$, from which the return maneuver impulse $\Delta v_B$ can then be calculated using Equation \eqref{eq:deltavb}. 

To optimize the accessible volume for fixed $\theta_A$ and $\Delta v_{B,\mathrm{target}}$, we define the cost function:

\begin{equation}
    \mathcal{J}(\mathbf{x}) = \left( \Delta v_B(\mathbf{x}) - \Delta v_{B,\mathrm{target}} \right)^2 - \, \mathcal{I}(\mathbf{x}),
\end{equation}

Where $\mathbf{x} = \{\phi_A, \Delta v_A, t_B\}$ and

\begin{equation}
    \mathcal{I}(\mathbf{x}) \!\!=\!\!
\begin{cases}
\displaystyle \!\int_{t_i}^{t_f}\!\!\!\!\! n(\mathbf{r}_1'(t))D(t|\, \mathbf{x})  v_1(t)  dt,\!\!\!\! & \text{if } \left| \!\dfrac{\Delta v_B(\mathbf{x})}{\Delta v_{B,\mathrm{target}}} \!- \!1 \right| \! < \! \varepsilon, \\
0, & \text{otherwise},
\end{cases}
\end{equation}

The trajectory deviation is:

\begin{dmath}
D(t|\theta_A,\phi_A,\Delta v_A, t_B) = \\
\begin{cases}
\lVert \mathbf{r}_2(t|\theta_A,\phi_A,\Delta v_A)-\mathbf{r}_1(t) \rVert & \text{if } t_A<t \leq t_{B} \\
\lVert \mathbf{r}_3(t|\theta_A,\phi_A,\Delta v_A, t_B)-\mathbf{r}_1(t) \rVert
& \text{if } t_{B}<t<t_C \\
\end{cases}
\label{eq:Dpre}
\end{dmath}

The integral $\mathcal{I}(\mathbf{x})$ quantifies the accessible volume along Lucy’s trajectory during the encounter window, from start time $t_i$ to end time $t_f$ (see Section~\ref{subsec:windows}). It is weighted by the spatial distribution of $L_5$ Jupiter Trojans along the nominal path, given by $n(\mathbf{r}_1’(t))$ and computed using Equations~\eqref{eq:rot}, \eqref{eq:r1}, and \eqref{eq:nnorm}. The optimization algorithm restricts solutions to those where the computed return maneuver $\Delta v_B(\mathbf{x})$ lies within $\varepsilon = 0.1\%$ of the target $\Delta v_{B,\mathrm{target}}$. The speed along the nominal trajectory is given by $v_1(t) = \lVert \mathbf{v}_1(t) \rVert$ (Equation~\eqref{eq:v1}).


For the pre-Patroclus encounter window, we ran a differential evolution optimization algorithm across a range of azimuthal angles, $\theta_A$, from $0^\circ$ to $360^\circ$ in steps of $30^\circ$. For each value of $\theta_A$, we explored return maneuver targets $\Delta v_{B,\mathrm{target}}$ of 5, 20, 35, and 50 m/s. The direction of the initial impulse maneuver was constrained to lie within $\phi_A \in [0, \pi]$. Based on the toy model, which showed that the impulse at EGA3 is a fraction of the return impulse, we set the free parameter $\Delta v_A$ to vary within the range $[0, \Delta v_{B,\mathrm{target}}]$. Finally, because the toy model suggested that the optimal return time occurs $\!\sim\!100$ days after the start of the pre-Patroclus window, we used bounds for the return maneuver time of $t_B \in [t_i - 60\,\mathrm{d},\, t_f]$.

\begin{figure}
    \centering
    \includegraphics[width=1\linewidth]{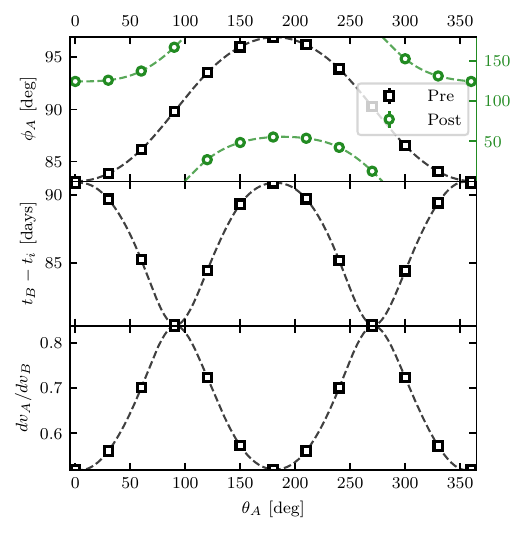}
    \caption{Parameters defining the accessible volume during Lucy’s pre- and post-Patroclus encounter windows. \textbf{Top:} Optimal polar angle $\phi_A$ as a function of azimuthal angle $\theta_A$. \textbf{Middle:} Optimal return time $t_B$ for the pre-Patroclus case, shown as a function of $\theta_A$. \textbf{Bottom:} Linear scaling between the initial maneuver magnitude $\Delta v_A$ and the return impulse $\Delta v_B$ for the pre-Patroclus trajectory. Black denotes the pre-Patroclus case, while green corresponds to the post-Patroclus scenario. Dashed lines indicate spline interpolations.}

    \label{fig:volumeparams}
\end{figure}

The results of the trajectory optimization for the pre-Patroclus encounter are shown in black in Figure~\ref{fig:volumeparams}. The first and second panels show the polar angle and return time (relative to the start of the window) for each azimuthal angle, respectively, averaged over all $\Delta v_{B,\mathrm{target}}$ values considered. The error bars, which are too small to be visible, indicate negligible variation with $\Delta v$, suggesting that both the return time and polar direction are effectively independent of the impulse magnitude. We confirmed that the relationship between the initial (EGA3) and return impulses is approximately linear. This enables estimating the required initial impulse as $\Delta v_A = (dv_A / dv_B) \Delta v_B$ with the scaling indicated in the third panel of Figure~\ref{fig:volumeparams}.

The optimization of the accessible volume for the post-Patroclus encounter window follows a similar approach to the pre-Patroclus case but is notably simpler. Here, $t_A$ corresponds to the start time of the post-Patroclus encounter window (Section~\ref{subsec:windows}). The accessible volume is constrained solely by the $\Delta v$ impulse applied at $t_A$ to initiate the deviation from the nominal trajectory. Since there are no constraints requiring Lucy to return to a specific target after the Patroclus encounter, there are no equivalent times $t_B$ or $t_C$ in this case. As a result, the only free parameter for the optimization is $\mathbf{x} = {\phi_A}$, given a fixed $\theta_A$ and $\Delta v_A$. The optimization procedure remains conceptually similar, with the cost function maximizing the accessible volume weighted by the spatial density of Trojans. Assuming $t_A$ marks the beginning of the encounter window, the nominal Lucy trajectory $\mathbf{r}_1(t)$ is calculated via Equation~\eqref{eq:r1}, and the modified trajectory $\mathbf{r}_2(t|\theta_A,\phi_A,\Delta v_A)$ is given by Equation~\eqref{eq:r2}. The resulting separation between the nominal and perturbed paths during the post-Patroclus window is therefore expressed as:

\begin{equation}
D(t|\theta_A,\phi_A,\Delta v_A) =
\lVert \mathbf{r}_2(t|\theta_A,\phi_A,\Delta v_A)-\mathbf{r}_1(t) \rVert
\label{eq:Dpost}
\end{equation}

The result of the optimization process for the post-Patroclus encounter window is shown in green in the first panel of Figure~\ref{fig:volumeparams}. As with the pre-Patroclus case, the polar angle was found to be independent of the magnitude of the impulse.

\section{Expected number of Trojans: Numerical approach} 
\label{app:numerical}

\begin{figure*}
    \centering
    \begin{minipage}{\textwidth}
        \centering
        \includegraphics[width=\linewidth]{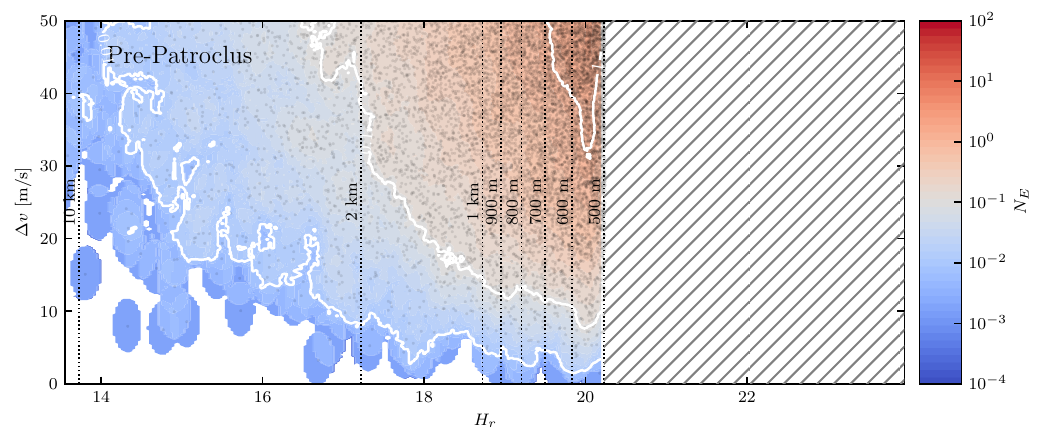}   
    \end{minipage}  
    \caption{Same as the top panel of Figure~\ref{fig:NE}, but derived from the numerical Monte Carlo simulation of encounter outcomes. Black semi-transparent dots represent the measured encounters from all 5000 trials. Hatched lines indicate the size range omitted from the simulation, as we are focused on the largest possible flyby targets and aimed to keep the simulation numerically tractable. Only the $N_E = 1$, 0.1, and 0.01 contours are shown. The colormap limits are identical to those in Figure~\ref{fig:NE}.}    
    \label{fig:numerical}
\end{figure*}

To validate our semi-analytical approach, we performed a numerical Monte Carlo evaluation of the expected number of encounterable Trojans in the pre-Patroclus case. This method involves generating multiple realizations of the $L_5$ cloud and propagating their orbits to assess the probability of encounter with Lucy.

To generate a synthetic realization of the Trojan cloud down to a given size threshold, we use inverse transform sampling. For a magnitude limit $H_{\text{lim}}$, we first compute the cumulative number of objects $\Sigma_{\text{lim}} = \Sigma(<H_{\text{lim}})$, then draw $\Sigma_{\text{lim}}$ samples from a uniform distribution $u \in [0,1]$. These are mapped to magnitudes using the following piecewise function:

\begin{dmath}
H(u) = \\ 
\begin{cases}
H_0+\frac{1}{\alpha_1} \log_{10}(u \Sigma_{\text{lim}}), & \text{if } 0\leq u < u_{B1} \\
\begin{split}
H_{B1} + \frac{1}{\alpha_2}\log_{10}(u \Sigma_{\text{lim}}) \\-\frac{\alpha_1}{\alpha_2}  (H_{B1} - H_0), \end{split} & \text{if } u_{B1} \leq u < u_{B2} \\
\begin{split}
H_{B2} + \frac{1}{\alpha_3} \log_{10}(u \Sigma_{\text{lim}}) \\ - \frac{\alpha_2}{\alpha_3} (H_{B2} - H_{B1}) \\ - \frac{\alpha_1}{\alpha_3} (H_{B1} - H_0) , \end{split} & \text{if } u_{B2} \leq u \leq 1
\end{cases}
\end{dmath}

with transition points:

\begin{align}
u_{B1} &= \frac{1}{\Sigma_{\text{lim}}}  10^{\alpha_1 (H_{B1} - H_0)} \\
u_{B2} &= \frac{1}{\Sigma_{\text{lim}}}  10^{\alpha_2 (H_{B2} - H_{B1}) + \alpha_1 (H_{B1} - H_0)}
\end{align}

This approach allows efficient sampling from the absolute magnitude distribution defined in Equation~\eqref{eq:hdistr}. Since our primary goal is to identify the largest possible flyby target while keeping the population size within computational limits, we set $H_{\text{lim}}$ in each realization to correspond to $D_{\text{lim}}=$ 500 m (assuming a 5$\%$ albedo), yielding on the order of half a million samples. For each sampled $H$, we assign a state vector using the spherical orbital basis as described in Section~\ref{sec:search}. The spatial coordinates are drawn from the $L_5$ spatial distribution (Section~\ref{subsec:spatialdistr}) at the start of the pre-Patroclus encounter window (Section~\ref{subsec:windows}). Because we generate multiple realizations of the $L_5$ cloud, we do not evaluate the dynamical stability of individual objects, which is an acceptable simplification given the short integration times and the low contamination shown in our previous tests.

Each orbit is propagated daily through the encounter window. At each epoch, we approximate the required $\Delta v$ to intercept the object as $\Delta v (t) = \Delta r(t)/|t - t_m|$, where $\Delta r(t)$ is the separation between Lucy and the Trojan, and $t_m$ is the time of the Patroclus encounter. For each orbit, we extract the minimum $\Delta v$. This process is repeated over multiple Monte Carlo realizations. The number of objects satisfying $\Delta v \leq \Delta v_{\text{max}}$ and $H_r \leq H_{r,\text{max}}$ yields a numerical estimate of the expected number of encounterable Trojans up to the given absolute magnitude and $\Delta v$ budget.

Figure~\ref{fig:numerical} shows the results of 5000 Monte Carlo trials. We combined the encounter distances from all simulations into a single KDE, producing a density map directly comparable to the semi-analytical results shown in Figure~\ref{fig:NE}. The region of parameter space most relevant to our study is where at least one object is expected ($N_E=1$). Notably, in both Figure~\ref{fig:NE} and Figure~\ref{fig:numerical}, the $N_E=1$ contour intersects the maximum $\Delta v$ of 50 m/s at an absolute magnitude of $H_r\sim19.5$ ($D\sim700\,$m). Similarly, the 0.1 and 0.01 density contours in Figure~\ref{fig:numerical} intersect the $\Delta v=50\,$m/s axis at the same absolute magnitudes as predicted analytically in Figure~\ref{fig:NE}. 

Despite the simplifications in this numerical test (finite timesteps, a limited number of trials, a linear $\Delta v$ approximation, no stability filter, and a population truncated to $D\ge500\,$m) the results are in strong agreement with the semi-analytical predictions. The only notable deviation is a slight edge effect on the $N_E=1$ contour around 500–550 m, attributed to both the truncation of the population at $D=500\,$m and the finite width of the kernel. Overall, this numerical experiment validates our semi-analytical approach and the results presented in Section~\ref{subsec:integral}.


\bibliography{main}{}
\bibliographystyle{aasjournal}


\end{document}